\documentclass[preprint]{elsarticle} 

\makeatletter
	\def\ps@pprintTitle{%
 	\let\@oddhead\@empty
	\let\@evenhead\@empty
	\def\@oddfoot{\centerline{\thepage}}%
	\let\@evenfoot\@oddfoot}
\makeatother

\usepackage{etoolbox}
\patchcmd{\MaketitleBox}{\footnotesize\itshape\elsaddress\par\vskip36pt}{\footnotesize\itshape\elsaddress\par\parbox[b][36pt]{\linewidth}{\vfill\hfill\textnormal{\today}\hfill\null\vfill}}{}{}%
\patchcmd{\pprintMaketitle}{\footnotesize\itshape\elsaddress\par\vskip36pt}{\footnotesize\itshape\elsaddress\par\parbox[b][36pt]{\linewidth}{\vfill\hfill\textnormal{\today}\hfill\null\vfill}}{}{}%

\usepackage[colorlinks=true,%
            breaklinks=true,%
            linkcolor=blue,
            urlcolor=blue,%
            citecolor=blue,%
            pdftitle={Title of paper}, 
            pdfkeywords={Keywords},	
            pdfauthor={Authors},		
            bookmarksopen=false,
            pdfpagemode=UseNone]{hyperref}

\usepackage[margin=0.9in]{geometry}
\usepackage{makecell}
\usepackage[T1]{fontenc}
\usepackage[english]{babel}
\usepackage{icomma}
\usepackage[latin1]{inputenc} 
\usepackage{subfiles}
\usepackage{xcolor}
\usepackage{tabularray}
\biboptions{numbers,sort&compress}
\usepackage[justification=centering]{subcaption}

\usepackage{amsmath}
\usepackage{pifont}

\usepackage{mathtools}
\usepackage{siunitx}
\usepackage{bbm}
\usepackage{amsmath}
\usepackage{amsfonts}
\usepackage{amsthm}
\usepackage{amssymb}
\usepackage{array}
\usepackage{algorithm} 
\usepackage{algorithmicx}
\usepackage{algpseudocode}
\usepackage{siunitx}
\usepackage{amsmath}
\usepackage{mathrsfs}
\usepackage{xpatch}

\usepackage{color}
\usepackage{url}
\usepackage{cleveref}
\usepackage{graphicx}
\usepackage{breakcites}
\usepackage{soul} 
\usepackage{enumitem}
\usepackage[title,titletoc,toc]{appendix}

\newtheoremstyle{mytheoremstyle}{5pt}{5pt}{\itshape}{}{\bfseries}{.}{.5em}{} 
\theoremstyle{mytheoremstyle}

\newtheoremstyle{myremarkstyle}{3pt}{3pt}{\itshape}{}{\bfseries}{.}{.5em}{} 
\theoremstyle{myremarkstyle}

\usepackage{pifont}

\usepackage[textsize=small]{todonotes}

\newcommand{\Hquad}{\hspace{0.5em}}

\begin{document}
\begin{frontmatter}
    \title{Secondary drift-driven instabilities in the presence of a parallel-propagating electromagnetic ion cyclotron wave and cold multi-component ions}
    \author[1,2]{Opal Issan\corref{cor1}}\ead{oissan@ucsd.edu}
    \author[3,4]{Patrick Kilian}
    \author[2,3]{Vadim Roytershteyn}
    \author[2]{Salomon Janhunen}
    \author[2]{Gian Luca Delzanno}

    \cortext[cor1]{Corresponding author}
    \address[1]{Department of Mechanical and Aerospace Engineering, University of California San Diego, La Jolla, CA, USA}
    \address[2]{T-5 Applied Mathematics and Plasma Physics Group, Los Alamos National Laboratory, Los Alamos, NM, USA}
    \address[3]{Space Science Institute, Boulder, CO, USA}
    \address[4]{Space Physics Research Group, Department of Physics, University of Helsinki, Helsinki, Finland}
    \begin{abstract}
    Electromagnetic ion cyclotron (EMIC) waves are commonly observed in Earth's inner magnetosphere, particularly during geomagnetic storms driven by anisotropic ring-current protons.
    While their role in radiation belt scattering of hot ions is well established, their interaction with the cold (less than 100 eV) plasma remains less understood. 
    This is partly due to limited magnetospheric cold ion observations, as spacecraft charging can prevent cold ions from reaching onboard instruments.
    It is well-known that the electric field of a parallel-propagating EMIC wave can drive inter-species perpendicular polarization drifts that excite lower-hybrid secondary instabilities. 
    In multi-component plasmas, these include the modified two-stream and the ion-ion cross-field instabilities. 
    In this paper, we study the impact of such secondary instabilities on the parallel-propagating EMIC wave and multi-component plasma via a fully kinetic particle-in-cell simulation and linear theory.
    We find that the secondary waves persist even at low EMIC amplitudes, provided the cold population remains sufficiently cold.
    The kinetic simulation demonstrates that these secondary modes produce anisotropic heating of cold protons and singly-charged oxygen ions, primarily in the direction perpendicular to the ambient magnetic field and of electrons in both parallel and perpendicular directions.
    \end{abstract}	
\end{frontmatter}

\section{Introduction}\label{sec:introduction}
Electromagnetic ion cyclotron (EMIC) waves are low-frequency electromagnetic fluctuations with wavelengths on the order of the ion inertial length and frequencies approaching the ion cyclotron frequency. 
EMIC waves typically propagate parallel or at slightly oblique angles to the ambient magnetic field and are left-hand polarized.
They are mainly generated in Earth's magnetosphere by anisotropic keV proton populations with a higher perpendicular than parallel temperature~\cite{gary_1976_jgr, gary_1993_theory}, namely the \textit{proton cyclotron anisotropy instability}.
The anisotropic protons are typically found in the ring current or plasma sheet and are a few percent of the total plasma density, where the rest of the populations consist of electrons, cold protons, and a small population of cold heavier ions~\cite{khazanov_2017_van_allen}. 
Here, \textit{cold} refers to energies below 100 eV. 
The magnetospheric cold populations are predominantly of ionospheric origin.
The presence of multiple ion populations allows the EMIC wave to occur in multiple frequency bands below the local ion cyclotron frequencies of different ion species, including protons (H$^{+}$)~\cite{kim_2025_emic_h+}, helium ions (He$^{+}$)~\cite{horne_1997_emic_he+}, oxygen ions (O$^{+}$)~\cite{xiongdong_2015_emic_o+}, and nitrogen ions (N$^{+}$)~\cite{bashir_2021_emic_n+}.
EMIC waves play a key role in magnetospheric dynamics by driving hot ion pitch-angle scattering, leading to the parallel acceleration and precipitation of ring current ions through cyclotron resonance~\cite{kennel_1966_jgr, cornwall_1970_jgr}. 
They are also resonant with relativistic radiation belt electrons, contributing to their precipitation~\cite{usanova_2014_emic, summers_2003_emic}.

Although EMIC waves are driven by anisotropic hot ions, their properties (frequency, wavenumber, and amplitude) are influenced by the presence of cold ions.
The cold ions act passively through their density since they are non-resonant with the EMIC wave; see~\citet[\S D]{cuperman_1981_anisotropy_review} and~\citet{gary_1994_jgr_emic} for a comprehensive review. 
As the EMIC wave reaches saturation, it can interact with cold ions through a range of mechanisms that are not yet fully understood in part due to limited \textit{in-situ} cold ion magnetospheric measurements. 
This is because spacecraft charging creates a positive potential barrier that prevents cold ions from reaching onboard detectors~\cite{delaznno_2021_cold_impact}. This barrier is typically a few volts in the plasmasphere but can reach tens of volts in the outer magnetosphere~\cite{maldonado_2023_frontiers}.

Nonlinear interactions between cold ions and parallel-propagating EMIC waves can lead to substantial energization over long timescales.
\citet{omidi_2010_phase_bunching} discovered using hybrid simulations that cold protons and singly-charged helium ions interacting with parallel-propagating EMIC waves undergo significant heating, which they attributed to a nonresonant nonlinear mechanism~\cite{mauk_1982_phase_bunching_grl, berchem_1985_phase_bunching, bortnik_2010_phase_bunching}. 
Furthermore, they found that the heating is predominantly perpendicular to the ambient magnetic field, with a weaker parallel component arising from nonlinearly generated electrostatic waves at half the EMIC wavelength~\cite{omidi_2010_phase_bunching}.
This process persists for $\sim10^{3}$ proton gyroperiods, after which the electrostatic structures dissipate; however, perpendicular ion heating continues on much longer timescales of $\sim10^{4}$ proton gyroperiods.
During the heating process, the primary EMIC wavepacket amplitude decreases by about 50\%. 
Similar hybrid simulations have been examined by~\cite{omura_1985_jgr, qian_1990_jgr, kwon_2023_jgr, kim_2024_jgr, gu_2025_jgr}, and analogous perpendicular energization of cold helium ions has been observed in the outer magnetosphere lobes~\cite{young_1981_jgr, roux_1982_jgr, anderson_1994_emic_jgr, kim_2024_jgr, abid_2021_pop}.

In addition to these long-timescale processes, cold ions can interact with EMIC waves through faster drift-driven secondary instabilities. These secondary instabilities occur on spatial scales comparable to the cold electron gyroradius and at frequencies in the lower-hybrid range~\cite{khazanov_1997_lh_emic}.
These instabilities are driven by the mass-dependent perpendicular polarization drift induced by the EMIC electric field.
\citet{sizonenko_1967_emic} and~\citet{khazanov_1996_lh_emic} developed the linear theory of such secondary instabilities in single-ion electron plasma.
The theory was later extended to a multi-component plasma by~\citet{khazanov_1997_lh_emic}, demonstrating that the EMIC electric field amplitude required to excite lower-hybrid waves is significantly lower in a hydrogen plasma with a cold heavy ion component (e.g., oxygen) than in a pure hydrogen plasma. 
The instability threshold is lower due to the addition of an ion-ion cross-field instability~\cite{gary_1987_drift, papadopoulos_1971_pof, barbosa_1986_pof} between the heavy ions and protons, on top of the modified two-stream instability~(MTSI)~\cite{ott_1972_mtsi, krall_1971_mtsi, mcbride_1972_mtsi,janhunen_2018_2Decdi} between electrons and ions.
Moreover, several observational studies have reported a correlation between lower-hybrid waves and EMIC waves, using data from the Van Allen Probes~\cite{khazanov_2017_van_allen}, Viking spacecraft~\cite{khazanov_1997_lh_observations, pottelette_1990_jgr}, and the Magnetospheric Multiscale mission~\cite{liu_2025_nature}.
These secondary drift-driven instabilities also occur in the presence of other low-frequency electromagnetic waves with perpendicular electric fields, such as whistler waves at ion scales~\cite{saito_2015_pop} and Alfv\'en waves~\cite{gamayunov_1992_alfven, singh_2007_jgr, khazanov_2007_grl}. Electron scale secondary drift-driven instabilities also occur in the presence of a parallel-propagating chorus whistler wave and cold plasma~\cite{roytershteyn_2021_pop, roytershteyn_2024_frontiers, issan_2026_pop}.

Here, we present a study to quantify the secondary drift-driven instabilities induced by EMIC waves using a fully kinetic nonlinear particle-in-cell~(PIC) simulation.
The 2D3V PIC setup consists of a homogeneous plasma composed of cold electrons~($\sim$eV) and three ion populations: a dense cold proton core~($\sim$eV), a less dense anisotropic hot proton population~($\sim$keV), and a less dense population of cold singly-charged oxygen ions~($\sim$eV).
The anisotropic hot protons excite the primary EMIC wave, which induces a polarization relative drift between the different species.
This configuration supports six secondary drift-driven instabilities, with three at high frequency (ion-ion acoustic, electron-ion acoustic, and electron-cyclotron drift) and three at lower frequency in the lower-hybrid range (an ion-ion cross-field instability and two electron-ion MTSIs)~\cite{gary_1987_drift}.
We focus on a parameter regime in which secondary lower-frequency instabilities dominate, thereby excluding higher-frequency modes.
In particular, we assume that the primary EMIC wave is low-amplitude, such that the perpendicular ion-to-electron relative drift is much smaller than the electron thermal speed, which suppresses beam cyclotron instabilities~\cite{buneman_1962_drift, sizonenko_1967_emic}, and that electron and cold ion temperatures are comparable, which prevents acoustic-like instabilities~\cite{sizonenko_1967_emic, gary_1987_drift}.
The presented 2D3V PIC simulation shows that the secondary lower-hybrid modes heat cold ions (protons and singly-charged oxygen) in the perpendicular direction and electrons in both the parallel and perpendicular direction.
Cold ion heating is primarily driven by a rapidly growing ion-ion cross-field instability, whereas electron heating is associated with a more slowly growing MTSI.
This leads to a redistribution of energy among the various plasma populations and introduces an additional pathway by which the parallel-propagating EMIC waves mediate energy transfer between hot and cold species.

This paper is organized as follows.
Section~\ref{sec:linear_secondary} presents the linear theory of the MTSI and ion-ion cross-field secondary instabilities driven by a parallel-propagating monochromatic EMIC wave.
Section~\ref{sec:PIC} provides a detailed analysis of the PIC simulations and compares the results with linear theory. 
Finally, section~\ref{sec:conclusion} presents the conclusions and outlines directions for future work.

\paragraph{Notation}{Consider a Cartesian coordinate system $\vec{x} = [x, y, z]^{\top}$. All quantities are in Gaussian units with a uniform magnetic field $\vec{B}_{0} = B_{0} \hat{z}$, $B_{0}>0$. The subscript `$s$' denotes the species. We consider four species: cold electrons (`$e$'), cold protons (`$pC$'), hot protons (`$pH$'), and singly-charged cold oxygen ions (`O$^{+}$'). The subscripts $\|$ and $\perp$ refer to the vector projections parallel and perpendicular to $\vec{B}_{0}$, respectively. The plasma is assumed to be quasi-neutral.
We define the cyclotron frequency $\Omega_{cs} \coloneqq q_{s} B_{0}/m_{s}c$, so that $\Omega_{ce}$ is negative.
We denote the thermal velocity as $v_{ts} \coloneqq \sqrt{T_{s}/m_{s}}$, plasma frequency as $\omega_{ps} \coloneqq \sqrt{4\pi n_{s} q_{s}^2/m_{s}}$, skin depth as $d_{s} \coloneqq c/\omega_{ps}$, plasma beta $\beta_{s} \coloneqq 8\pi n_{s}T_{s}/B_{0}^2$, gyroradius $\rho_{s} \coloneqq v_{ts}/|\Omega_{cs}|$, lower-hybrid frequency $\omega_{LH} = \sqrt{\omega_{pp}^2/[1 + \omega_{pe}^2/\Omega_{ce}^2]} \approx \sqrt{\Omega_{cp}|\Omega_{ce}|}$, and anisotropy level as $A_{s} \coloneqq T_{\perp s}/T_{\| s} - 1$, where $c$ is the speed of light in a vacuum, $e$ is the positive elementary charge, and $q_{s}, m_{s}, T_{s}, n_{s}$ are the charge, rest mass, temperature, and density of species $s$, respectively. 
The concentration ratio of each species is denoted by $c_{s} \coloneqq n_{s}/n_{e}$. 
We refer to the parallel-propagating EMIC wave as the ``primary'' or ``driver'' wave. 
The wave-normal angle, i.e., the angle between the wavevector $\vec{k}$ and the background magnetic field $\vec{B}_{0}$, is denoted by $\theta_{k}$. 
The velocity vector cylindrical coordinate system is denoted by $\vec{v} \coloneqq [v_{\perp}\cos(\theta_{v}) \Hquad v_{\perp} \sin(\theta_{v}) \Hquad v_{\|}]^{\top}$, where $\theta_{v}$ is the gyro-phase angle. 
We denote the Fourier transform of an arbitrary function $f(\vec{x}, t)$ by $\hat{f}(\vec{k}, \omega) \coloneqq \int \mathrm{d}^{3} x \int \mathrm{d} t f(\vec{x}, t) \exp(-i \vec{k}\cdot \vec{x} + i\omega t)$, where $\omega \coloneqq \omega_{r} + i \gamma$ is the complex frequency with real, oscillatory component $\omega_r$ and growth rate $\gamma$.  
}

\section{Linear theory of secondary electrostatic drift-driven instabilities}\label{sec:linear_secondary}
\begin{figure}
    \centering
    \includegraphics[width=0.6\linewidth]{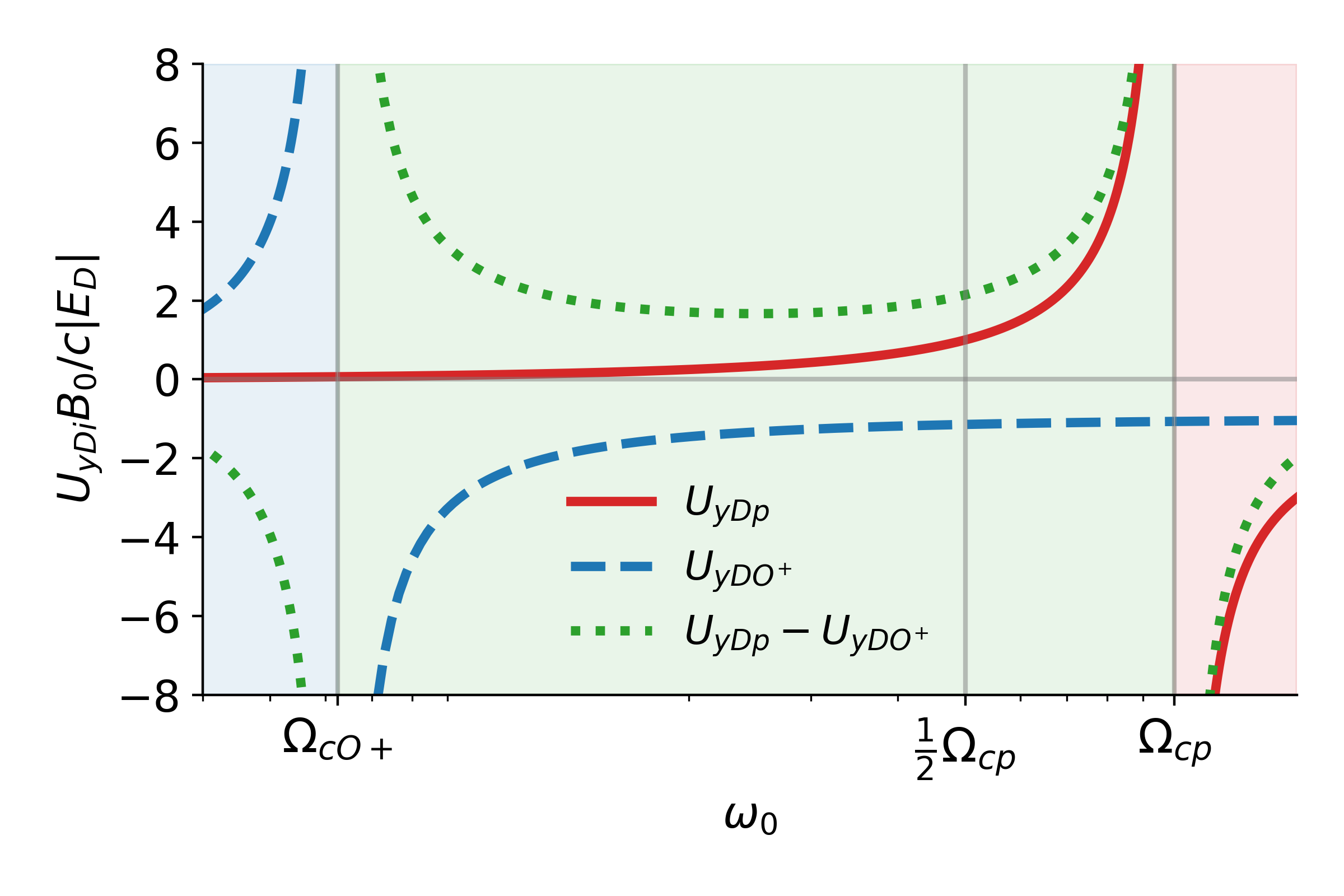}
    \caption{The relative drift of each ion species (proton and oxygen) at $t=z=0$  with respect to electrons (and each other), see Eq.~\eqref{relative_drift_y}. The drift direction reverses for different ion species depending on the primary wave frequency $\omega_{0}$, which leads to secondary waves propagating in opposite perpendicular directions.}
    \label{fig:drift_estimates_vs_omega0}
\end{figure}

A plasma subject to an alternating electric field perpendicular to the background magnetic field, generates a transverse relative drift between different plasma species~\cite[\S 6.2.3]{akhiezer_1975_book}. 
Here, the relative drift is driven by a monochromatic parallel-propagating EMIC wave, which is left-hand polarized and characterized by the following electric field
\begin{equation}\label{EMIC_E_field}
    \vec{E}_{D}(z, t) \coloneqq  |E_{D}| \left[\cos(\omega_{0} t - k_{\|0}z), \Hquad -\sin(\omega_{0} t- k_{\|0}z), \Hquad 0\right]^{\top},
\end{equation}
where $\omega_{0} \sim \Omega_{cp} \ll |\Omega_{ce}|$ is the driver frequency, $k_{\|0}$ is the driver parallel wavenumber, and $|E_{D}|$ is the driver amplitude. 
The equations of motion for the polarization drift $\vec{V}_{Ds}(z, t)$ driven by the electric field of the EMIC wave in Eq.~\eqref{EMIC_E_field} are 
\begin{equation}\label{dVdt}
    \frac{\mathrm{d} \vec{V}_{Ds}(z, t)}{\mathrm{d} t} = \frac{q_{s}}{m_{s}} \left[\vec{E}_{D}(z, t) + \frac{\vec{V}_{Ds}(z, t) \times \vec{B}_{0}}{c}\right].
\end{equation}
Let $V_{\perp Ds}(t) \coloneqq V_{xDs}(t) + i V_{yDs}(t)$, such that Eq.~\eqref{dVdt} yields
\begin{align*}
    \frac{\mathrm{d}V_{\perp Ds}(z, t)}{\mathrm{d} t} &= \frac{q_{s}}{m_{s}}\left[|E_{D}|\exp(-i[\omega_{0} - k_{\|0}z]) - \frac{i B_{0}V_{\perp Ds}(z, t)}{c} \right].
\end{align*}
We assume the solution is of the form $V_{\perp Ds}(z, t) \propto \exp(-i[\omega_{0} t - k_{\|0}z])$, which results in
\begin{equation*}
    V_{\perp Ds}(z, t) = i \frac{q_{s}}{m_{s}}\frac{|E_{D}|}{\omega_{0} - \Omega_{cs}} \exp(-i[\omega_{0} t - k_{\|0}z]).
\end{equation*}
Since the drift is mass-dependent, there is a relative drift between the different particle populations. 
We assume the primary wave frequency is near the ion cyclotron frequency, i.e. $\omega_{0} \sim \Omega_{cp} \ll |\Omega_{ce}|$, such that the relative drift between ions and electrons in the $y$-direction is
\begin{equation}\label{relative_drift_y}
    U_{yDi}(z, t) \coloneqq V_{yDi}(z, t) - V_{yDe}(z, t) =  \frac{e \omega_{0}|E_{D}|}{m_{i} \Omega_{ci} [\Omega_{ci} - \omega_{0}]} \cos\left(\omega_{0}t - k_{\|0}z\right).
\end{equation}
Figure~\ref{fig:drift_estimates_vs_omega0} shows the relative drift in Eq.~\eqref{relative_drift_y} at $t=z=0$ for protons and singly-charged oxygen ions.
The proton-oxygen relative drift exceeds the proton-electron and oxygen-electron drift when $ \tfrac{1}{16}\Omega_{cp} = \Omega_{cO+} < \omega_{0} < \Omega_{cp} $ (away from the asymptotes). For $ \Omega_{cp}< \omega_{0} $, the proton-electron relative drift dominates, whereas for $\omega_{0} < \Omega_{cO+}$, the oxygen-electron relative drift dominates.
This suggests that the ion-ion cross-field instability has larger relative drifts in comparison to MTSI in the frequency range $\Omega_{cO+} < \omega_{0} < \Omega_{cp}$, and \textit{vice versa} outside this frequency range.
Additionally, if $\Omega_{cO+} < \omega_{0} < \Omega_{cp}$, the drift direction reverses for protons and oxygen, leading to secondary electron-proton MTSI and oxygen-electron MTSI modes propagating in opposite (positive and negative) perpendicular directions. 

From this point onward, we neglect the $z$-dependence of the driving EMIC field, since the secondary instabilities are dominated by short-wavelength fluctuations, which allows for a local approximation, e.g. $z=0$.
This is justified by the scale separation $k_{\|} \gg k_{\|0} \sim 1/d_{p}$, where $k_{\|0}$ and $k_{\|}$ denote the parallel wavenumbers of the primary EMIC wave and the secondary instabilities, respectively.
A transformation into the frame of reference co-drifting with the electron population can be obtained by a change of phase space coordinates $(\vec{x}, \vec{v}) \to (\vec{x}', \vec{v}')$:
\begin{equation}\label{co_drifting_frame}
    \vec{x}' \coloneqq \vec{x} - \int_{0}^{t} \vec{V}_{De}( \tau) \mathrm{d} \tau \qquad \mathrm{and} \qquad \vec{v}' \coloneqq \vec{v} - \vec{V}_{De}(t).
\end{equation} 
The cold electron electrostatic Vlasov-Poisson equation in the co-drifting frame~\eqref{co_drifting_frame} is
\begin{align}
    \left[\partial_{t}  + \vec{v}'  \cdot \nabla_{\vec{x}'}  - \frac{e}{m_{e}} \left[\delta \vec{E}(\vec{x}', t) + \frac{\vec{v}' \times \vec{B}_{0}}{c} \right] \cdot \nabla_{\vec{v}'} \right]f_{e}(\vec{x}', \vec{v}', t)&=0, \label{vlasov_cold_co_drifting}\\
    \nabla_{\vec{x}'} \cdot \delta \vec{E}(\vec{x}', t) \coloneqq -\nabla_{\vec{x}'}^{2}\delta \phi(\vec{x}', t) &= 4\pi \sum_{s} q_{s} \int \mathrm{d}^3 v' f_{s}(\vec{x}', \vec{v}', t),\label{poisson_equation}
\end{align}
where $\delta \vec{E}(\vec{x}', t)$ is the secondary electrostatic electric field and $\delta \phi(\vec{x}', t)$ is the electric potential, see~\ref{sec:appendix-codrift-electron} for a detailed derivation. 
Let $f_{s}(\vec{x}', \vec{v}', t) \coloneqq F_{0s}(\vec{v}') + \delta f_{s}(\vec{x}', \vec{v}', t)$, such that the equilibrium $F_{0s}(\vec{v}')$ is bi-Maxwellian and gyrotropic ($\partial_{\theta_{v}} F_{0s}= 0$):
\begin{equation}\label{bi_maxwellian}
    F_{0s}(\vec{v}') \coloneqq \frac{n_{s}}{\pi^{3/2} \alpha_{\perp s}^2 \alpha_{\| s}} \exp \left( - \frac{{v_{\|}'}^2}{\alpha_{\| s}^2} - \frac{{v_{\perp}'}^2}{\alpha_{\perp s}^2}  \right) ,\qquad
    \alpha_{\perp s}^{2} \coloneqq \frac{2 T_{\perp s}}{m_{s}}\qquad \mathrm{and} \qquad 
    \alpha_{\| s}^{2} \coloneqq \frac{2 T_{\| s}}{m_{s}}. 
\end{equation} 
After linearizing Eq.~\eqref{vlasov_cold_co_drifting}, the resulting electron linear response in a magnetized plasma with a bi-Maxwellian equilibrium~\eqref{bi_maxwellian} in the co-drifting frame~\eqref{co_drifting_frame} is
\begin{align}
    \delta \hat{n}_{e}(\vec{k}, \omega) &\coloneqq \int  \mathrm{d}^3 v' \int  \mathrm{d}^{3} x' \int \mathrm{d} t  \delta f_{e}(\vec{x}', \vec{v}', t) \exp(-i \vec{k}\cdot \vec{x}' + i\omega t) \nonumber\\
    &= \frac{e n_{e}}{T_{\| e}}  \delta \hat{\phi}(\vec{k}, \omega) \left[ 1 +   \sum_{n=-\infty}^{\infty} \Gamma_{n}(\lambda_{e}) Z(\xi_{e}^{n}) \left[\xi_{e}^{0} +  \frac{n |\Omega_{ce}|}{k_{\|} \alpha_{\| e}} \left[ 1-\frac{\alpha_{\| e}^2}{\alpha_{\perp e}^2} \right]\right]\right], \label{electron_response_co_drifting}
\end{align}
where $\omega \coloneqq \omega_{r} + i \gamma$ is the frequency,  $\vec{k} = [0, \Hquad k_{\perp}, \Hquad k_{\|}]^{\top}$ is the wavevector, $Z$ is the plasma dispersion function~\cite{fried_conte_1961}, $\Gamma_{n}(\lambda_{e}) \coloneqq \exp(-\lambda_{e}) I_{n}(\lambda_{e})$, $I_{n}$ is the modified Bessel function of the first kind, $\lambda_{e} \coloneqq k_{\perp}^2 \alpha_{\perp e}^2/2\Omega_{ce}^2$, $\xi_{e}^{n} \coloneqq [\omega - n|\Omega_{ce}|]/|k_{\|}| \alpha_{\| e}$, see~\citet[\S 2]{gary_1993_theory} for a detailed derivation of the electrostatic electron response in magnetized plasma.

Since we assume that $\omega_{0} \ll \omega \sim \omega_{LH}$, the relative ion drift $\vec{U}_{Ds}$ can be treated as time-independent, allowing a Galilean transformation into the drift frame.
The unmagnetized ion response with a bi-Maxwellian equilibrium~\eqref{bi_maxwellian} in the co-drifting frame is
\begin{equation}\label{ion_response_co_drifting}
   \delta \hat{n}_{s}(\vec{k}, \omega) \coloneqq \int  \mathrm{d}^3 v' \int  \mathrm{d}^{3} x' \int \mathrm{d} t \delta f_{s}(\vec{x}', \vec{v}', t) \exp(-i \vec{k}\cdot \vec{x}' + i\omega t)  =  \frac{q_{s} n_{s}}{2T_{\| s}} \delta \hat{\phi}(\vec{k}, \omega) Z'\left(\frac{\omega - \vec{k} \cdot \vec{U}_{Ds}}{\sqrt{k_{\perp}^2 \alpha_{\perp s}^2 + k_{\|}^2\alpha_{\|s}^2}}\right).
\end{equation}
Inserting the electron~\eqref{electron_response_co_drifting} and ion~\eqref{ion_response_co_drifting} response in the co-drifting frame in the Poisson equation~\eqref{poisson_equation} results in 
\begin{align}
    |\vec{k}|^2+ \frac{2\omega_{pe}^2}{\alpha_{\|e}^2} 
    &\left[ 1 +   \sum_{n=-\infty}^{\infty} \Gamma_{n}(\lambda_{e}) Z(\xi_{e}^{n}) \left[\xi_{e}^{0} +  \frac{n |\Omega_{ce}|}{k_{\|} \alpha_{\| e}} \left[ 1-\frac{\alpha_{\| e}^2}{\alpha_{\perp e}^2} \right]\right]\right]\label{dispersion_relation}\\
    &= \sum_{s\in\{O^{+}, pH, pC\}}\frac{\omega_{ps}^2}{\alpha_{\| s}^2 }  Z' \left( \frac{\omega - \vec{k} \cdot \vec{U}_{Ds}}{\sqrt{k_{\perp}^2 \alpha_{\perp s}^2 + k_{\|}^2\alpha_{\|s}^2}}\right).\nonumber
\end{align}
To obtain analytic estimates for the secondary instabilities, it is useful to examine the cold-plasma limit of Eq.~\eqref{dispersion_relation}. 
The large argument asymptotic expansion of the plasma dispersion function is given by~\cite[\S 3]{hunana_2019_closure}:
\begin{equation}\label{analytic_response_asymptotics_high}
    Z(\xi) =  -\frac{1}{\xi}  - \frac{1}{2\xi^3} + \mathcal{O}\left(\frac{1}{\xi^{5}}\right)\qquad \mathrm{for} \qquad |\xi| \gg 1 \qquad \mathrm{and} \qquad \text{Im}(\xi) >0.
\end{equation}
Since we assume that $\omega \sim \omega_{LH} \ll |\Omega_{ce}|$, it follows that $|\xi_{e}^{n}| \gg 1$ for $n \neq 0$. Under this condition, Eq.~\eqref{electron_response_co_drifting} can be simplified using the large-argument expansion of the plasma dispersion function in Eq.~\eqref{analytic_response_asymptotics_high}, together with the assumption of isotropic electrons $T_{\|e} \approx T_{\perp e}$, yielding
\begin{equation*}
    \delta \hat{n}_{e}(\vec{k}, \omega) = \frac{e n_{e}}{T_{\| e}}  \delta \hat{\phi}(\vec{k}, \omega) 
    \left[ 1 +  \Gamma_{0}(\lambda_{e}) Z(\xi_{e}^{0}) \xi_{e}^{0} \right].
\end{equation*}
Also, if ions are isotropic and cold, i.e. $|\omega - k_{\perp} U_{yDs}| \geq \sqrt{2} v_{ts} |\vec{k}| $, $|\xi_{e}^{0}| \gg 1$, and $\lambda_{e} \ll 1$ with $\Gamma_{0}(\lambda_{e}) \approx 1 - \lambda_{e}$, then Eq.~\eqref{dispersion_relation} simplifies to 
\begin{align}\label{dispersion_relation_cold}
    1 - \frac{\omega_{pe}^2}{\omega^2 } \cos^2(\theta_{k}) + \frac{\omega_{pe}^2}{\Omega_{ce}^2} \sin^2(\theta_{k})
    = \sum_{s\in \{O^{+}, pH, pC\}}\frac{\omega_{ps}^2}{[\omega - k_{\perp} U_{yDs}]^2}.
\end{align}

\paragraph{Modified two-stream instability}{
We consider a two-species ion-electron plasma and assume an isothermal plasma with $T_e \sim T_i$ and a drift ordering $U_{yDi} \sim v_{ti} \ll v_{te}$, such that the dispersion relation~\eqref{dispersion_relation_cold} reduces to
\begin{equation}\label{dispersion_cold_mtsi}
    1 - \frac{m_{i}}{m_{e}c_{i}} \frac{\tilde{\omega}_{LH}^2}{\omega^2 }\cos^2(\theta_{k})
    - \frac{\tilde{\omega}_{LH}^2}{[\omega - k_{\perp} U_{yDi}]^2} = 0, \qquad \tilde{\omega}_{LH} \coloneqq \sqrt{\frac{\omega_{pi}^2}{1 + \omega_{pe}^2/\Omega_{ce}^2}}.
\end{equation}
We look for unstable modes near $\omega \approx k_{\perp} U_{yDi}+ \eta$ with $\eta \ll k_{\perp} U_{yDi} $, in which case Eq.~\eqref{dispersion_cold_mtsi} becomes
\begin{equation*}
    1 -  \frac{m_{i}}{m_{e}c_{i}}\frac{\tilde{\omega}_{LH}^2}{k_{\perp}^2 U_{yDi}^{2}}\left[ 1 - \frac{2 \eta}{k_{\perp} U_{yDi}} \right] \cos^2(\theta_{k}) 
    - \frac{\tilde{\omega}_{LH}^2}{\eta^2} = 0
\end{equation*}
Then, if the coefficient of the quadratic term is small, i.e. $\tilde{\omega}_{LH} \approx k_{\perp} U_{yDi}$ and $\cos^{2}(\theta_{k}) \approx c_{i}m_{e}/m_{i}$, we get
%
\begin{equation*}
   \omega =  k_{\perp} U_{yDi} + \sqrt[3]{1}\left[\frac{1}{2} \frac{ m_{e}c_{i}}{m_{i}}\frac{1}{\cos^{2}(\theta_{k})}\right]^{\frac{1}{3}} \tilde{\omega}_{LH}, \qquad \mathrm{where} \qquad
     \sqrt[3]{1} = \left(1, \frac{-1 + \sqrt{3}i}{2}, \frac{-1 - \sqrt{3} i}{2}\right).
\end{equation*}
The growth and frequency of the unstable MTSI branch are  
\begin{equation}\label{MTSI_estimate}
    \gamma = \frac{\sqrt{3}}{2^{\frac{4}{3}}} \left[\frac{m_{e}c_{i}}{m_{i}}\frac{1}{ \cos^2(\theta_{k})}\right]^{\frac{1}{3}} \tilde{\omega}_{LH} \qquad \mathrm{and} \qquad \omega_{r} = \left[1 - \frac{1}{2^{\frac{4}{3}}} \left[\frac{m_{e}c_{i}}{m_{i}} \frac{1}{\cos^2(\theta_{k})}\right]^{\frac{1}{3}} \right] \tilde{\omega}_{LH}.
\end{equation}
Note that the estimate above is for oblique modes and is not valid for strictly perpendicular wave-normal angles. 
The MTSI growth rate has a weak dependence on the ion-to-electron mass ratio $\gamma/\Omega_{ci} \propto [m_{i}/m_{e}]^{1/6}$. 
In section~\ref{sec:PIC}, we employ a PIC simulation with a reduced mass ratio $m_{p}/m_{e} = 400$; under a realistic proton-to-electron mass ratio, the MTSI growth rate is therefore expected to be $1.28$ times larger.
Moreover, the unstable modes are oblique with $k_{\|}/|\vec{k}| = \cos(\theta_{k}) \sim \sqrt{c_{i} m_{e}/m_{i}}$ and marginal stability is achieved when $v_{ti} \approx U_{yDi}$~\cite{ott_1972_mtsi}.
}

\paragraph{Ion-ion cross-field instability}{Consider an electron-proton-oxygen plasma, the cold limit of the dispersion relation~\eqref{dispersion_relation_cold} with $k_{\|} = 0$ reduces to 
\begin{equation*}
    1 - \frac{\omega_{LH}^2 c_{pC}}{\omega^2} - \frac{c_{O+}m_{p}\omega_{LH}^2}{m_{O+}\left[\omega - k_{\perp} [U_{yDO+} -  U_{yDpC}]\right]^2} = 0.
\end{equation*}
If $\omega = k_{\perp} [U_{yDO+} - U_{yDpC}] + \eta$ and $\eta \ll 1$, then
\begin{equation*}
    1 - \frac{\omega_{LH}^2 c_{pC}}{k_{\perp}^2 [U_{yDO+} - U_{yDpC}]^2} \left[1 - \frac{2\eta}{k_{\perp} [U_{yDO+} - U_{yDpC}] } \right]- \frac{c_{O+}m_{p}\omega_{LH}^2}{m_{O+} \eta^2} = 0.
\end{equation*}
Additionally, if $k_{\perp}^2 [U_{yDO+} - U_{yDpC}]^2 \approx \omega_{LH}^2 c_{pC}$, the growth and frequency of the ion-ion cross-field unstable branch are 
\begin{equation*}
    \gamma = \frac{\sqrt{3c_{pC}}}{2^{\frac{4}{3}}} \left[\frac{c_{O+}}{c_{pC}} \frac{m_{p}}{ m_{O+}} \right]^{\frac{1}{3}} \omega_{LH} \qquad \mathrm{and} \qquad \omega_{r} = \sqrt{c_{pC}}\left[1 - \frac{1}{2^{\frac{4}{3}}} \left[\frac{c_{O+}}{c_{pC}} \frac{m_{p}}{ m_{O+}} \right]^{\frac{1}{3}} \right] \omega_{LH} .
\end{equation*}
Increasing the oxygen and cold-proton density enhances the growth rate as
$\gamma \propto c_{O+}^{1/3}c_{pC}^{1/6}$. Moreover, the growth rate dependence on the mass ratio is stronger than MTSI, i.e. $\gamma/\Omega_{cp} \propto [m_{p}/m_{O+}]^{1/3} [m_{p}/m_{e}]^{1/2}$. Thus, using the reduced proton-to-electron mass ratio in section~\ref{sec:PIC} preferentially increases the growth rate of the ion-ion cross-field instability relative to the MTSI (Eq.~\eqref{MTSI_estimate}), reflecting its stronger $[m_p/m_e]^{1/2}$ scaling compared to the $[m_p/m_e]^{1/6}$ scaling of the MTSI. In particular, the reduced mass ratio used in section~\ref{sec:PIC} is $m_{p}/m_{e} = 400$, whereas the realistic value $m_{p}/m_{e} = 1836$ would yield an ion-ion cross-field growth rate approximately $2.14$ times larger.
}

\begin{figure}
    \centering
    \noindent\includegraphics[width=0.6\textwidth]{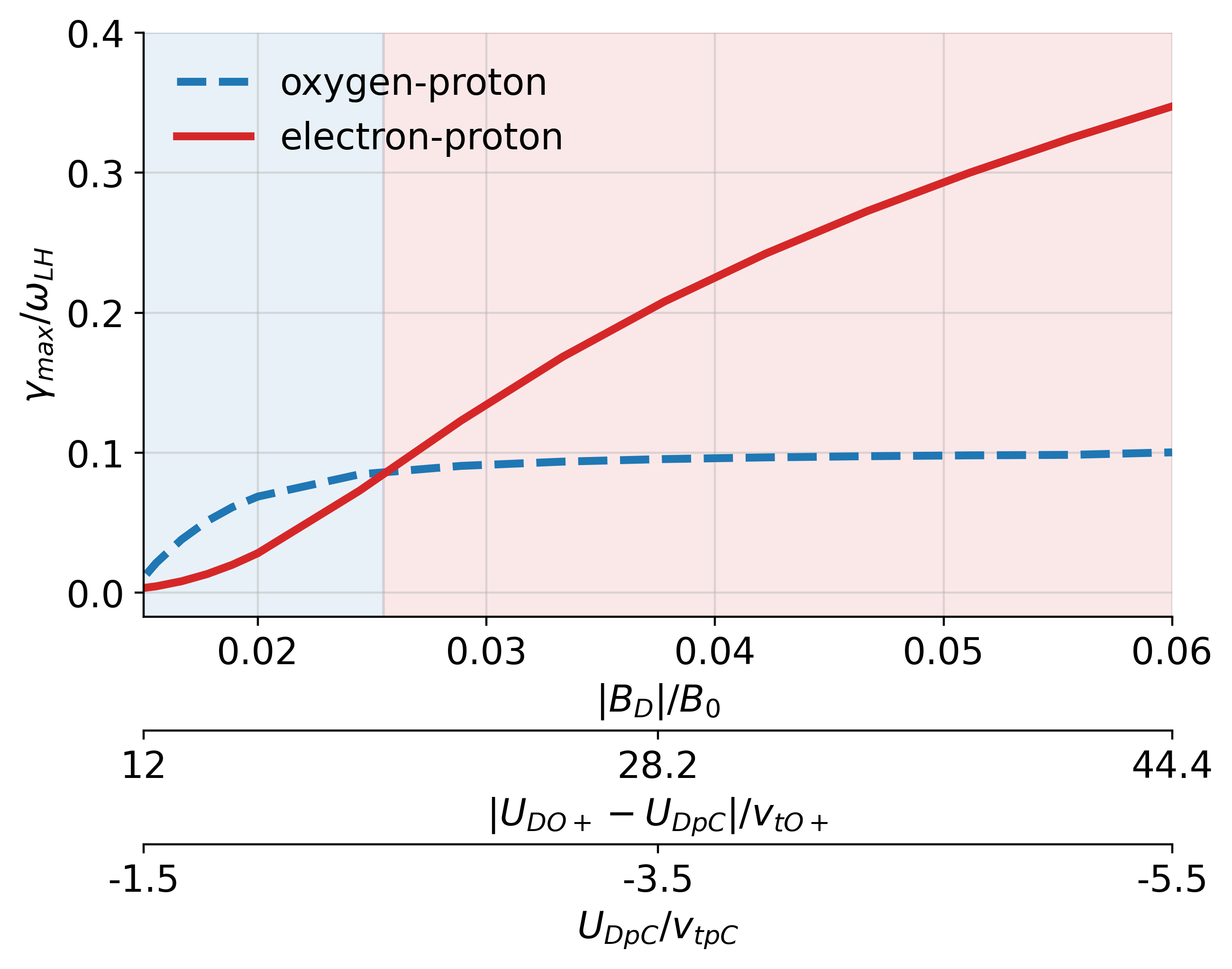}
    \caption{Dependence of the maximum growth rate of the secondary instabilities on the amplitude of the EMIC driver.}
    \label{fig:driver_amplitude_dependence}
\end{figure}

Figure~\ref{fig:driver_amplitude_dependence} illustrates how the secondary instabilities depend on the amplitude of the primary EMIC wave. The parameters are consistent with the PIC simulations presented later in section~\ref{sec:PIC}, with $\omega_{pe}/|\Omega_{ce}| = 2$, $\beta_{e} = 10^{-4}$, $m_p/m_e = 400$, $m_{O^+}/m_{p} = 16$, $T_{e} = T_{pC} = T_{O^+}$, $T_{\|pH} = 10^{4} T_{e}$, and $T_{\perp pH} = 4 T_{\|pH}$. The driver frequency is set to $\omega_{0} = 0.5\Omega_{cp}$. The relative drifts are shown on the additional horizontal axis of Figure~\ref{fig:driver_amplitude_dependence}.
From Eq.~\eqref{relative_drift_y}, the amplitude of the perpendicular drift scales linearly with the EMIC electric field amplitude $|E_D|$. 
Using Faraday's law, the corresponding magnetic field amplitude is given by $|B_D| = \frac{c k_{\|0}}{\omega_0} |E_D|$, where $|B_D|$ denotes the EMIC magnetic field amplitude. Therefore, the perpendicular drifts are also proportional to the EMIC magnetic field amplitude~$|B_D|$.
The numerical results of the dispersion relation in Figure~\ref{fig:driver_amplitude_dependence} indicate a clear transition in the dominant instability mechanism for the considered parameter regime.
Below the threshold amplitude $|B_D|/B_0 \approx 0.025$, the ion-ion cross-field instability dominates, whereas above this threshold the electron-proton MTSI becomes the dominant mode.
The growth rate of the ion-ion cross-field instability saturates as the EMIC wave amplitude increases.
Therefore, the most unstable drift-driven secondary instability depends both on the frequency and amplitude of the primary EMIC wave, as shown in Figures~\ref{fig:drift_estimates_vs_omega0} and~\ref{fig:driver_amplitude_dependence} respectively.

\section{Particle-in-cell simulation}\label{sec:PIC}
We outline the PIC simulation setup in section~\ref{sec:pic_setup}, and present the simulation results together with a comparison to linear theory in section~\ref{sec:pic_numerical_results}.

\subsection{Particle-in-cell simulation setup}\label{sec:pic_setup}
We study the secondary drift-driven instabilities via a fully kinetic nonlinear 2D3V PIC simulation using the Vector Particle-In-Cell (VPIC) code~\cite{bowers_2008_VPIC}. 
In the PIC simulation, EMIC waves are excited by a temperature anisotropy instability associated with a tenuous energetic population of hot protons and dense cold populations.
Specifically, we consider a four component plasma that is initialized with three cold populations (cold electrons with density $n_{e}$, cold protons of density $n_{pC} = 0.8 n_{e}$, and cold oxygen with density $n_{O+} = 0.1 n_{e}$) and a hot proton population with density $n_{pH} = 0.1 n_{e}$.
The cold populations are initialized with isotropic Maxwellian distributions of temperature $T_{e} = T_{pC} = T_{O+} \approx 6.38$ eV.
The hot proton population is initialized with a bi-Maxwellian~\eqref{bi_maxwellian} distribution with parallel temperature $T_{\| pH}=10^4 T_{e}$, and perpendicular temperature $T_{\perp pH} = 4 T_{\|pH}$. Thus, the hot proton anisotropy is initially $A_{pH}\coloneqq T_{\perp pH}/T_{\|pH} -1 = 3$. 
The initially uniform background magnetic field is $\vec{B}_0 = B_{0} \hat{z}$ and $\beta_{e} \coloneqq 8 \pi n_{e} T_{e}/B_{0}^2 = 10^{-4}$.
Other parameters are $\omega_{pe}/|\Omega_{ce}|=2$, reduced proton-to-electron mass ratio $m_p/m_e = 400$, and realistic oxygen-to-proton mass ratio $m_{O+}/m_p=16$.
The simulation is performed in a periodic domain with perpendicular length of $L_{y} \approx 2.87 d_{e} \approx 0.14 d_{p}$ and parallel length of $L_{z} \approx 232.71 d_{e} \approx 11.64 d_{p}$. 
The spatial resolution is $n_{y} \times n_{z} = 200 \times 16,000$ cells, which corresponds to a cell size of approximately $4$ cold proton Debye length. 
This is marginal at best for an explicit PIC code like VPIC, but is required to obtain a simulation of a reasonable computational cost.
The main effect of under-resolving the Debye length is substantial numerical energization observed in the simulations, which, for this system, appears not to impact the results significantly.
We set the time step to $\Delta t \omega_{pe} \approx 10^{-2}$, and each population is initially represented by $n_{ppc} = 4,000$ computational particles per cell. 

\begin{figure}
    \centering
    \begin{subfigure}{0.495\textwidth}
    \centering
    \caption{Primary instability frequency at $t=0$}
    \includegraphics[width=\linewidth]{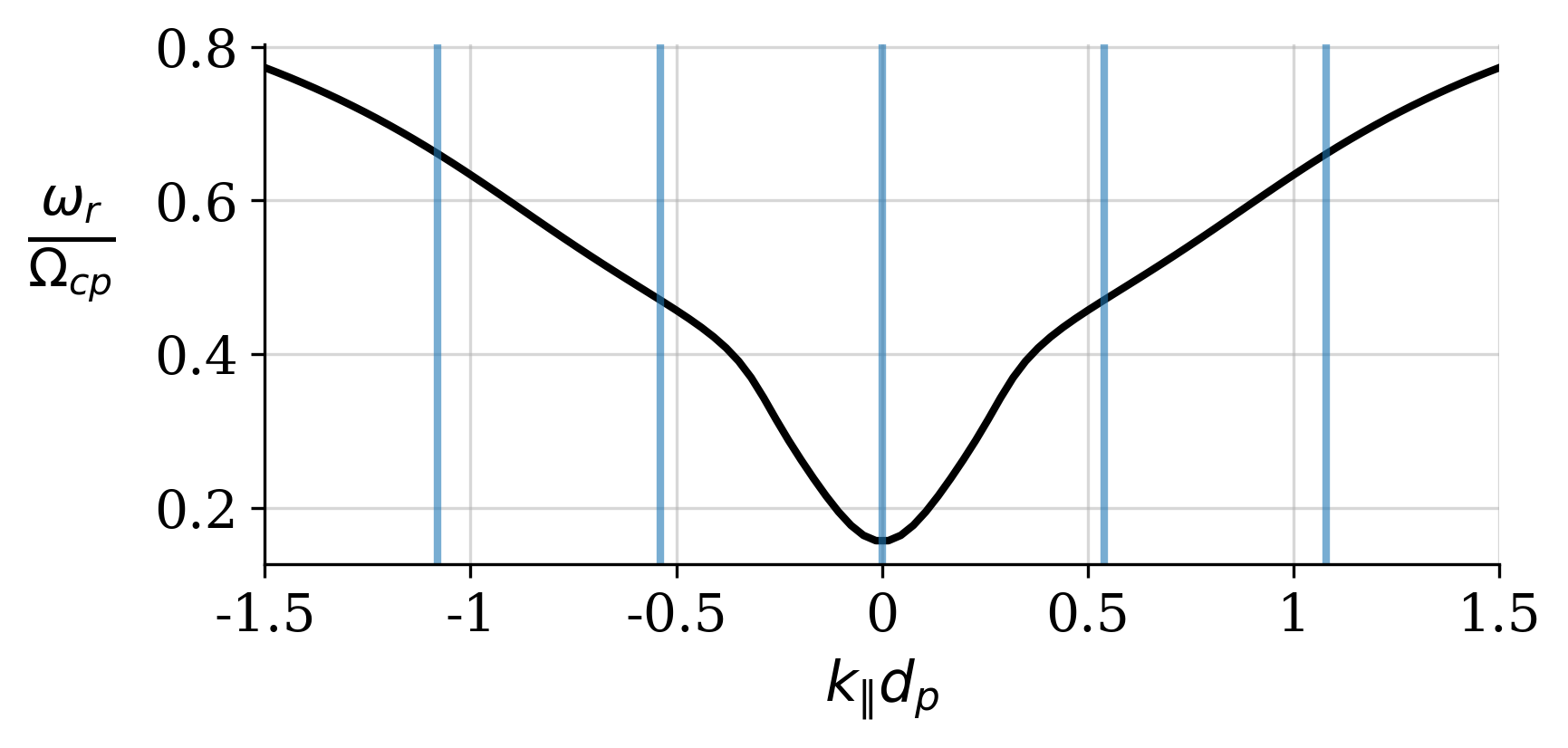}
    \end{subfigure}
    \begin{subfigure}{0.495\textwidth}
    \centering
    \caption{Primary instability growth rate at $t=0$}
    \includegraphics[width=\linewidth]{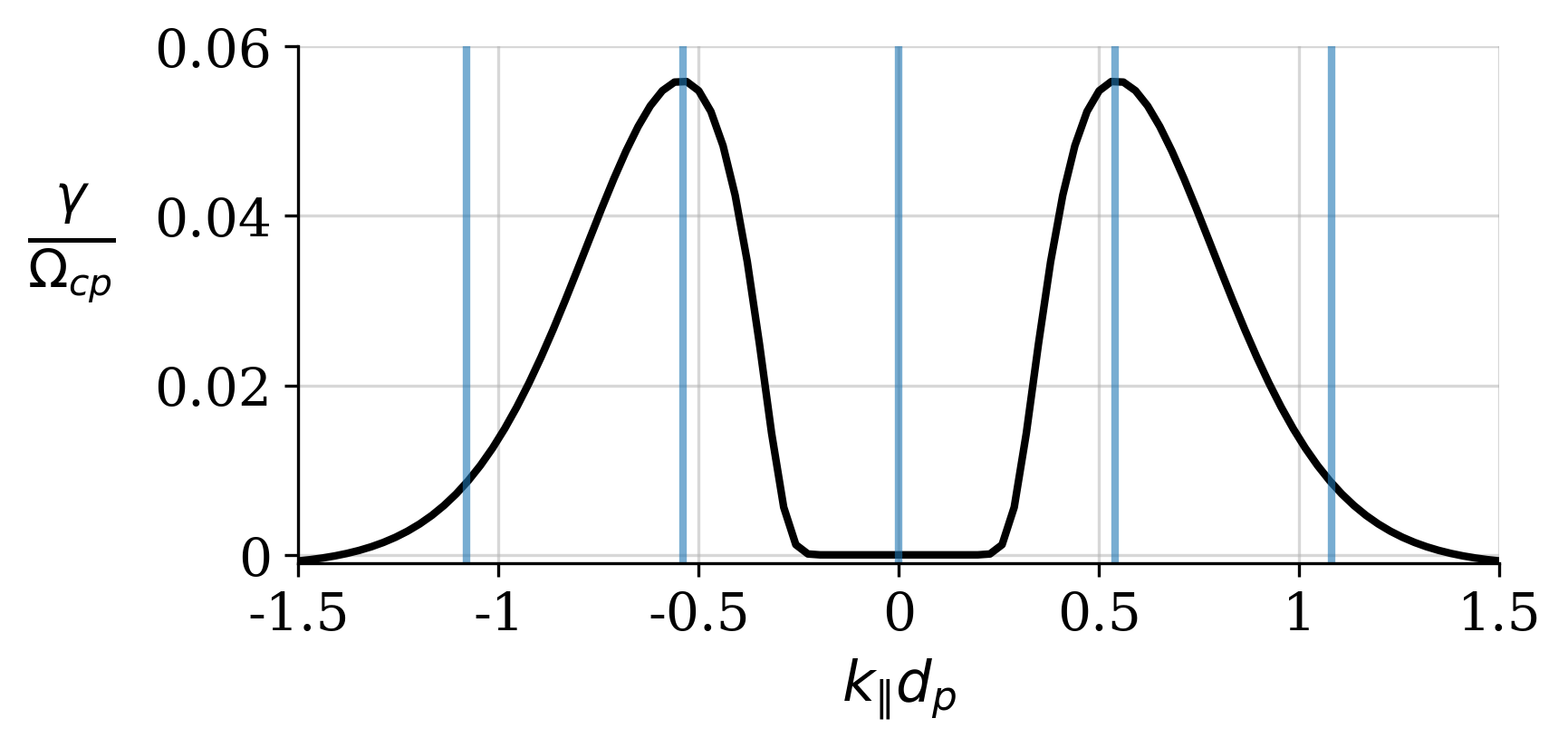}
    \end{subfigure}
    \caption{The primary proton cyclotron anisotropy instability (a) frequency and (b) growth rate at $t=0$. The vertical blue lines represent the PIC simulation resolution, such that only a few unstable modes are captured in the simulation, with the most unstable mode with $\omega_{0} \approx 0.5 \Omega_{cp}$, $|k_{\|0}|d_{p} \approx 0.5$, and $\gamma_{0} \approx 0.055 \Omega_{cp}$.}
    \label{fig:primary_instability_dispersion_relation}
\end{figure}

The primary proton cyclotron anisotropy instability dispersion relation for bi-Maxwellian equilibrium~\eqref{bi_maxwellian} is
\begin{equation}\label{primary_dispersion_relation}
    \sum_{s}\omega_{ps}^2 \left[\xi_{s}^{0} Z(\xi_{s}^{-1}) + A_{s} [1 + \xi_{s}^{-1} Z(\xi_{s}^{-1})]\right]= k_{\|}^2c^2 - \omega^2 \qquad \mathrm{where} \qquad \xi_{s}^{n} \coloneqq \frac{\omega + n \Omega_{cs}}{|k_{\|}| \alpha_{\| s}},
\end{equation}
see~\citet[\S 7.2.2]{gary_1993_theory} for a detailed derivation. 
Eq.~\eqref{primary_dispersion_relation} is valid for the parallel-propagating limit $k_{\perp} = 0$.
Figure~\ref{fig:primary_instability_dispersion_relation} shows the numerical results of the primary instability dispersion relation in Eq.~\eqref{primary_dispersion_relation} with the PIC initial parameters.
The PIC simulation resolution (illustrated by vertical blue lines in Figure~\ref{fig:primary_instability_dispersion_relation}) can capture only a few unstable modes, yet the first mode with $|k_{\|0}|d_{p} \approx 0.5$ is significantly more unstable and will dominate the primary wave properties. Its corresponding frequency is $\omega_{0} \approx 0.5 \Omega_{cp}$,with a growth rate of $\gamma_{0} \approx 0.055 \Omega_{cp}$.
In the initialization, we perturb the uniform equilibrium by imposing a  helical perturbation of magnetic field $\delta B_x = \delta B_0 \cos (k_{\|} z )$ and $\delta B_y = \delta B_0 \sin (k_{\|} z )$ with amplitude $\delta B/B_0 \approx 2 \times 10^{-3}$ and wavenumber $k_{\|0} = 2 \pi/L_z \approx 0.5 d_p^{-1}$. The perturbation seeds the growth of the forward-propagating EMIC wave. As discussed in section~\ref{sec:linear_secondary} and shown in Figure~\ref{fig:drift_estimates_vs_omega0}, we expect that for the given EMIC frequency the oxygen-proton drift will dominate over the proton-electron and oxygen-electron drifts.

\subsection{Particle-in-cell Numerical Results}\label{sec:pic_numerical_results}
\begin{figure}
    \centering
    \includegraphics[width=\linewidth]{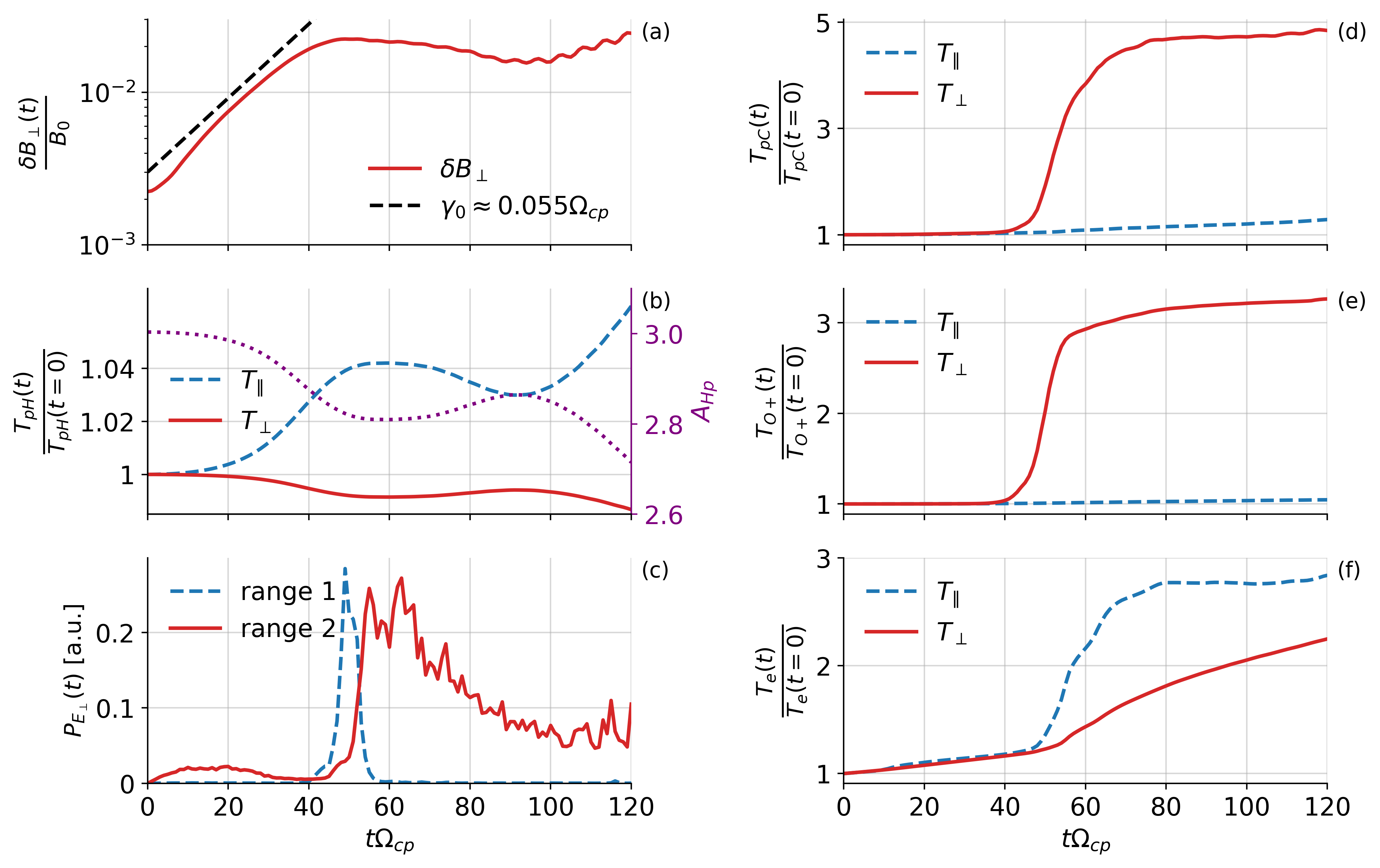}
    \caption{Time evolution of the PIC macroscopic quantities. Initially at $t\Omega_{cp} \in [0, 50]$, the proton cyclotron anisotropy instability leads to (a) growth of magnetic field perpendicular fluctuations and (b) a reduction in the hot proton anisotropy. Subfigure~(c) shows the electric power over different wavenumber ranges. ``Range 1" corresponds to $k_{\perp}\rho_{e} \in [0, 0.15]$ and $\theta_{k} \in [86^{\circ}, 90^{\circ}]$, and "range 2'' to $k_{\perp}\rho_{e} \in [0.15, 0.4]$ and $\theta_{k} \in [89.5^{\circ}, 90^{\circ}]$. The ``range 2" electric power is multiplied by 25. At $t \Omega_{cp} \approx 40$, the EMIC wave electric field excites a secondary ion-ion cross-field instability, resulting in predominantly perpendicular heating of (d) cold protons and (e) cold oxygen ions. Subfigure~(f) shows electron heating is in both the parallel and perpendicular directions due to a slowly growing MTSI. } 
    \label{fig:macro_quantities_tracers}
\end{figure}

Figure~\ref{fig:macro_quantities_tracers} summarizes the overall time evolution of the average macroscopic quantities in the 2D3V PIC simulation. 
Initially, the proton cyclotron anisotropy instability drives the growth of the perpendicular magnetic field fluctuations associated with the parallel-propagating EMIC wave, as shown in panel~(a). Meanwhile, the hot proton anisotropy, i.e. the source of free energy, decreases, as shown in panel~(b). 
The magnetic field perturbation growth rate agrees with linear theory (i.e., $\gamma_{0} \approx 0.055 \Omega_{cp}$).
Both the magnetic field perturbations and the hot proton temperature saturate at $t \Omega_{cp} \approx 50$. 
Panel~(c) shows the perpendicular electric power, which is defined as  $P_{E_{\perp}}(K, t) \coloneqq \sum_{\vec{k} \in K} |\hat{E}_{y}(\vec{k}, t)|^2$ for two ranges. 
``Range 2'' is set to  $k_{\perp}\rho_{e} \in [0, 0.15]$ and $\theta_{k} \in [86^{\circ}, 90^{\circ}]$. 
For visualization, the ``range 2'' electric power in panel~(c) is scaled up by a factor of $25$. 
The power in ``range 2'' decays much more slowly than in ``range 1''.
At $t\Omega_{cp} \approx 40$, the MTSI and ion-ion cross-field instability overlap in wavenumber space such that distinguishing their contribution to perpendicular electric power is challenging. They technically propagate in opposite perpendicular directions; however, because the simulation is periodically saved every $\Delta t \Omega_{cp} \approx 1$, these modes cannot be easily distinguished.
However, after $t\Omega_{cp} \approx 50$, the MTSI unstable modes exhibit longer wavelengths in comparison to the perpendicular ion-ion cross-field modes. Accordingly, we associate ``range 1'' primarily with ion-ion cross-field contributions and ``range 2'' predominantly with MTSI. 
The secondary ion-ion cross-field instability leads to predominantly perpendicular heating of cold protons ($\times 5$), see panel~(d), and cold oxygen ($\times 3$), see panel~(e). The most intense ion heating is observed during the time interval where the ``range 1'' perpendicular electric power peaks, as shown in panel~(c).  
The cold electrons in panel~(f) are heated in both the perpendicular ($\times 2$) and parallel ($\times 3$) directions. 
The electron heating is slower than ions due to slowly growing MTSI modes driven by electron-proton drift, as shown in the ``range 2'' electric power in panel~(c).
At the same time, as the secondary modes develop, the EMIC wave amplitude decreases from 
$\delta B_{\perp}/B_{0} \coloneqq \smash{[\delta B_{x}^2 + \delta B_{y}^2]^{1/2}}/2 \approx 0.022$ at $t\Omega_{cp} \approx 50$ to $\delta B_{\perp}/B_{0} \approx 0.015$ at $t\Omega_{cp} \approx 100$, corresponding to a 32\% reduction in wave amplitude. 
Then, the EMIC wave amplitude increases again with $\delta B/B_{0} \approx 0.024$ at $t\Omega_{cp} = 120$, corresponding to $60\%$ increase in wave amplitude in comparison to $t\Omega_{cp} = 100$. 
The hot proton anisotropy reduces from $A_{pH} = 3$ at $t=0$ to $A_{pH} = 2.84$ at $t\Omega_{cp} = 100$, and $A_{pH} = 2.71$  at $t\Omega_{cp} = 120$.

\begin{figure}
    \centering
    \begin{subfigure}{0.49\textwidth}
    \centering
    \caption{Frequency}
    \includegraphics[width=\linewidth]{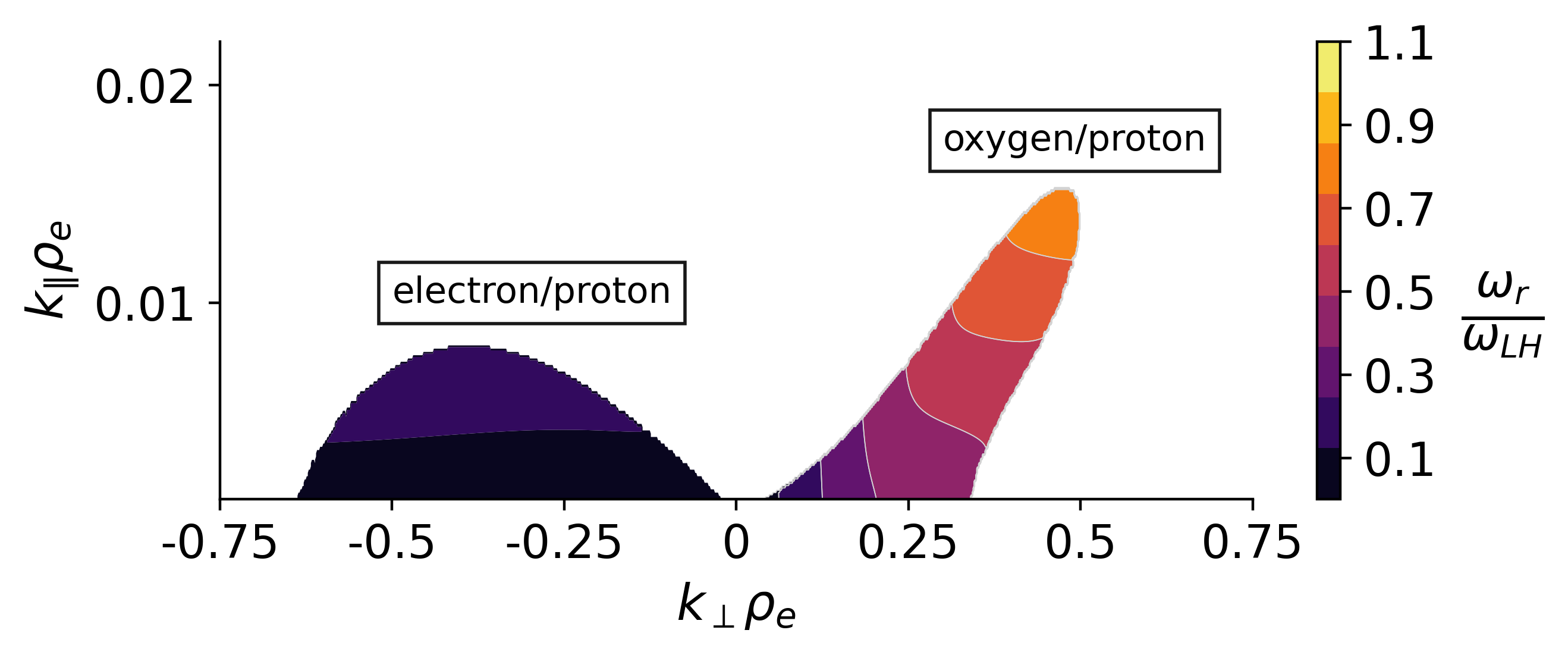}
    \end{subfigure}
    \begin{subfigure}{0.46\textwidth}
    \centering
    \caption{Growth rate}
    \includegraphics[width=\linewidth]{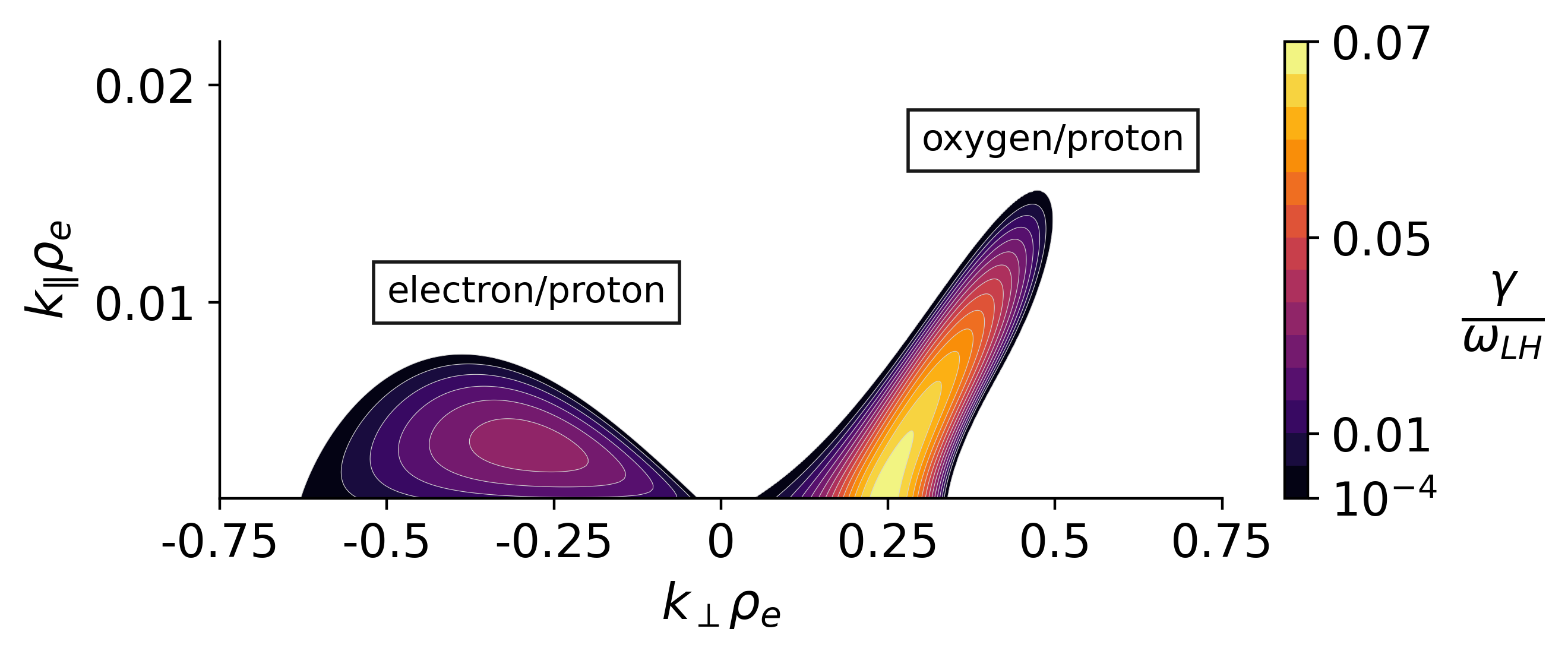}
    \label{fig:linear_theory_t_40_growth_rate}
    \end{subfigure}
    \caption{The secondary instabilities (a) frequency and (b) growth rate of the dispersion relation in Eq.~\eqref{dispersion_relation}. The fastest growing mode is attributed to the ion-ion cross-field instability driven by oxygen-proton drifts. }
    \label{fig:linear_theory_t_40}
\end{figure}

\begin{figure}
    \centering
    \begin{subfigure}{0.5\textwidth}
    \centering
    \caption{Oxygen-electron drift instability with $c_{pC} = 0$}
    \includegraphics[width=\linewidth]{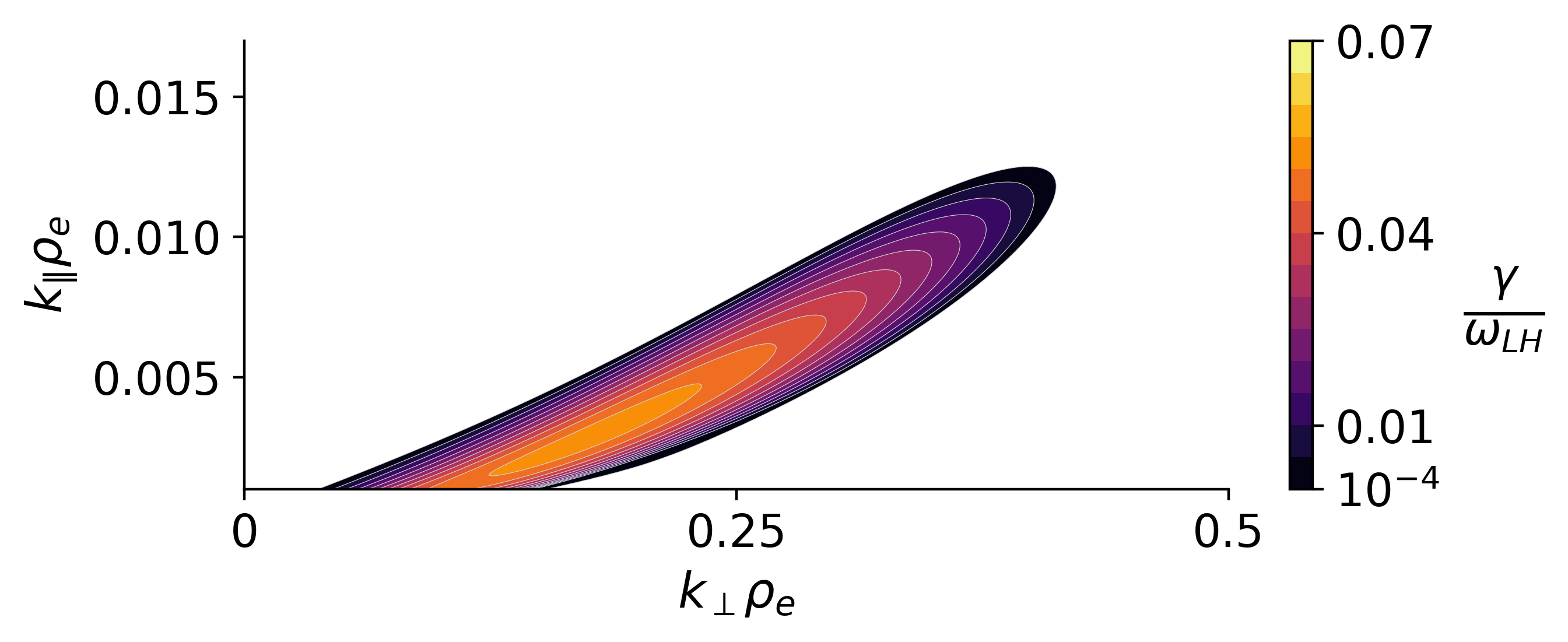}
    \end{subfigure}
    \hspace{-7pt}
    \begin{subfigure}{0.5\textwidth}
    \centering
    \caption{Oxygen-proton drift instability with $c_{pC}=0.8$}
    \includegraphics[width=\linewidth]{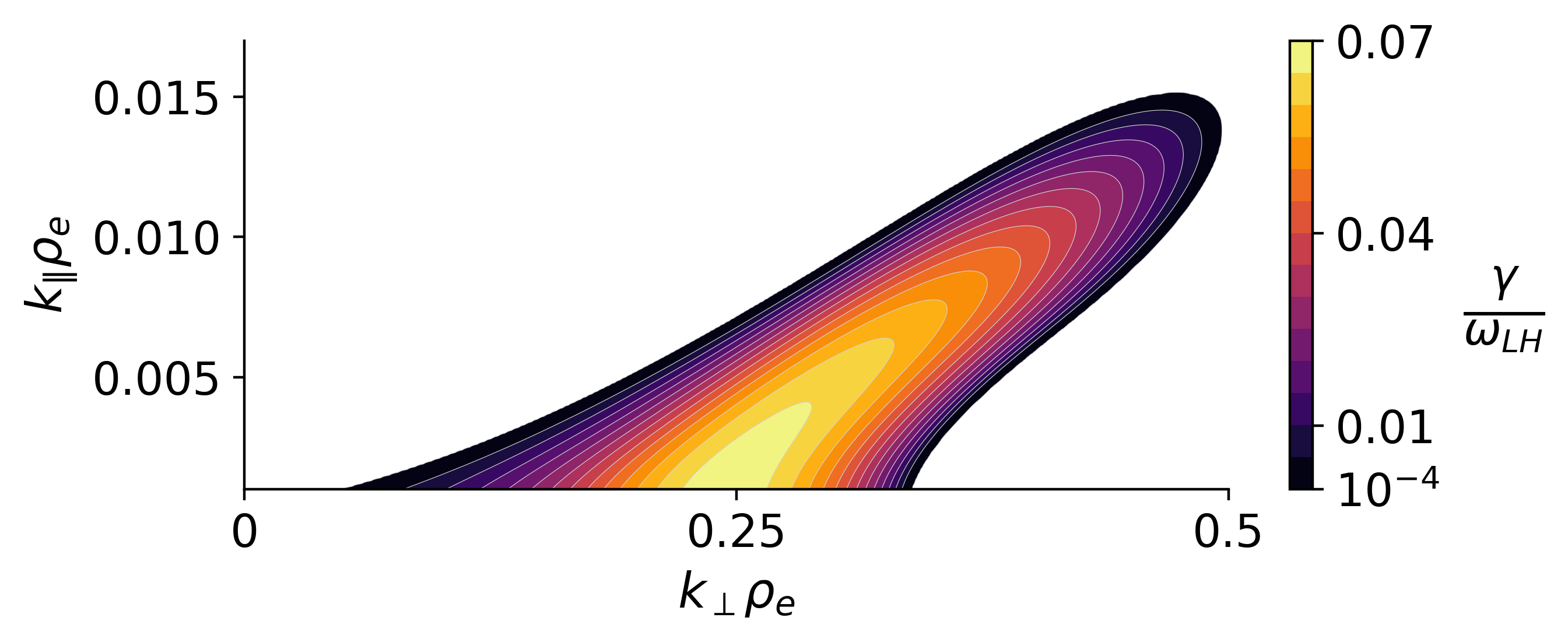}
    \end{subfigure}
    \caption{Secondary instabilities driven by (a) oxygen-electron drift ($c_{pC} = 0$) and (b) oxygen-proton drift ($c_{pC} = 0.8$). The unstable modes partially overlap in wavenumber space; however, the oxygen-proton instability exhibits a larger maximum growth rate, $\gamma_{\mathrm{max}} \approx 0.067\omega_{LH} \approx 1.34\Omega_{cp}$, and peaks at perpendicular propagation ($\theta_k \approx 90^\circ$). In contrast, the oxygen-electron instability is quasi-perpendicular ($\theta_k \approx 89^\circ$) with a smaller growth rate $\gamma_{\mathrm{max}}\approx 0.056\omega_{LH} \approx 1.12 \Omega_{cp}$. }
    \label{fig:linear_theory_electron_oxygen_vs_proton_oxygen}
\end{figure}

The numerical solution of the dispersion relation in Eq.~\eqref{dispersion_relation} for $t\Omega_{cp} \approx 40$ is shown in Figure~\ref{fig:linear_theory_t_40}.
The perpendicular relative drifts between the three ion species and the electrons, as defined in Eq.~\eqref{relative_drift_y} and obtained from the PIC simulation at $t \Omega_{cp} \approx 40$, are $U_{yDpC} \approx U_{yDpH} \approx -2 v_{tpC}$ and $U_{yDO+} \approx 8 v_{tO+}$. 
The parallel relative drifts are negligible $U_{zDpC} \approx U_{zDpH} \approx U_{zDO+} \approx 0$.
The remaining parameters are identical to those used in the PIC initialization, see section~\ref{sec:pic_setup}.
The ion-ion cross-field instability exhibits the largest growth rate with $\gamma_{\mathrm{max}} \approx 0.068\omega_{LH} \approx 1.36 \Omega_{cp}$ in the perpendicular direction. 
In comparison, the fastest-growing proton-electron MTSI mode is quasi-perpendicular with a wave-normal angle of $\theta_{k} \approx 89^{\circ}$ and a maximum growth rate of $\gamma_{\mathrm{max}} \approx 0.028\omega_{LH} \approx 0.56 \Omega_{cp}$. 
In addition, the dominant ion-ion cross-field mode has a higher real frequency, $\omega_r \approx 0.4\omega_{LH} \approx 8 \Omega_{cp}$, compared to $\omega_r \approx 0.1\omega_{LH} \approx 2 \Omega_{cp}$ for the proton-electron MTSI. 
Figure~\ref{fig:linear_theory_t_40} further shows that the unstable branches driven by electron-proton and oxygen-proton drifts propagate in opposite perpendicular directions, consistent with the opposing drift orientations. By contrast, the branches associated with oxygen-electron and oxygen-proton drifts propagate in the same perpendicular direction and exhibit overlap in wavenumber space.
This overlap is confirmed in Figure~\ref{fig:linear_theory_electron_oxygen_vs_proton_oxygen}, which shows the growth rate in wavenumber space for the oxygen-electron MTSI case with setting $c_{pC}=0$ and the oxygen-proton ion-ion cross-field instability case  with setting $c_{pC}=0.8$ (i.e., a zoomed in version of Figure~\ref{fig:linear_theory_t_40_growth_rate}). 
As expected, the oxygen-electron MTSI mode remains quasi-perpendicular, with $\theta_k \approx 89^{\circ}$ and $\gamma_{\mathrm{max}} \approx 0.052\omega_{LH} \approx 1.04\Omega_{cp}$. Nevertheless, the ion-ion cross-field instability remains dominant due to the larger effective relative drift, i.e. $|U_{yDO+} - U_{yDpC}|\approx 16v_{tO+}$ compared with $U_{yDO+} \approx 8v_{tO+}$, and low amplitude driver with $|B_{D}|/B_{0} \approx 0.02$.

\begin{figure}
    \centering
    \begin{subfigure}{0.5\textwidth}
    \centering
    \caption{Frequency}
    \includegraphics[width=\linewidth]{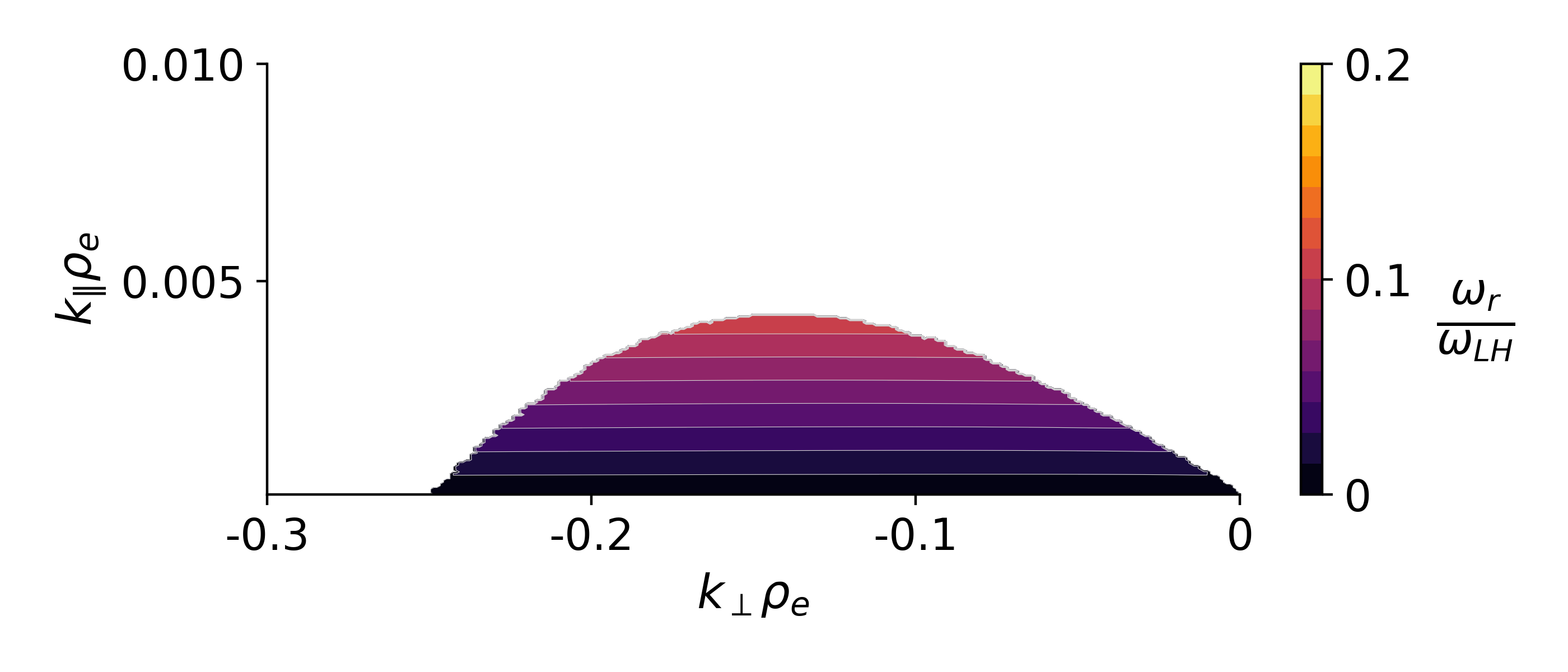}
    \end{subfigure}
    \hspace{-7pt}
    \begin{subfigure}{0.5\textwidth}
    \centering
    \caption{Growth rate}
    \includegraphics[width=\linewidth]{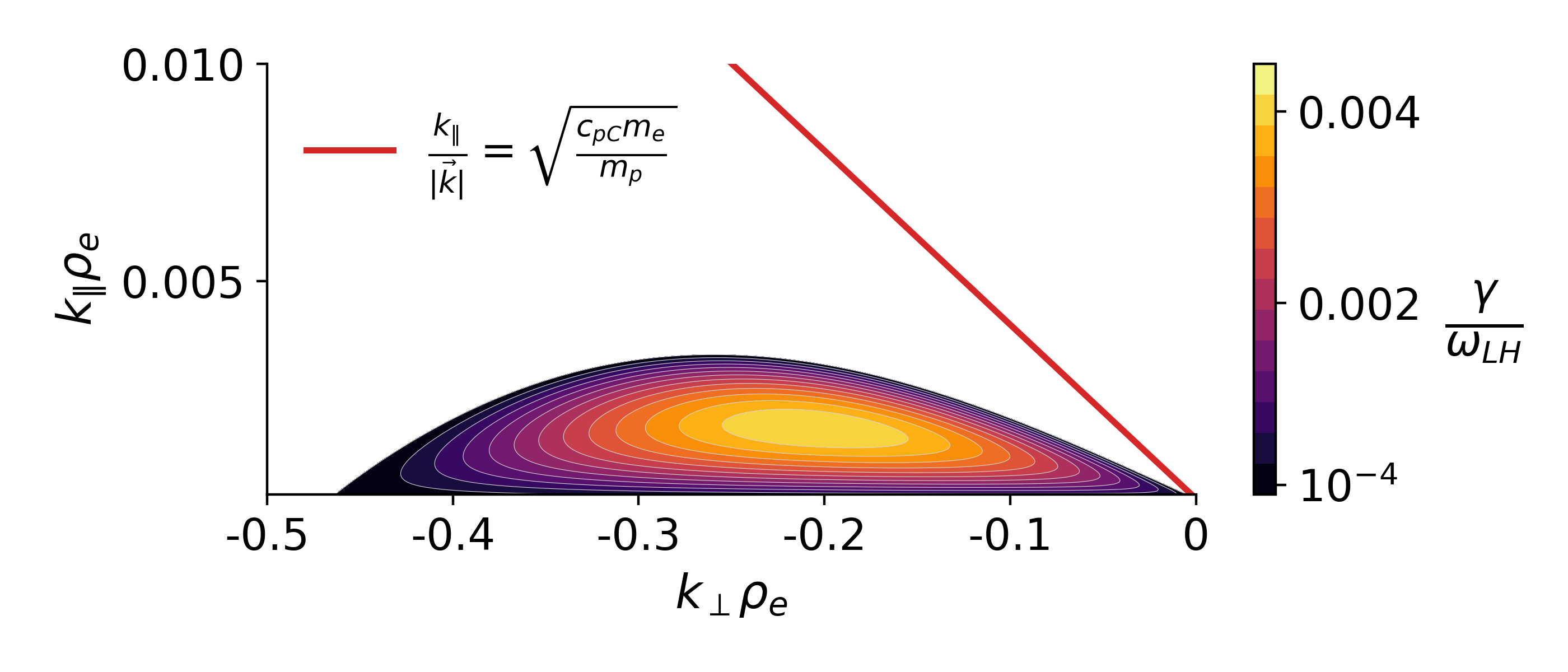}
    \end{subfigure}
    \caption{The electron-proton MTSI (a) frequency and (b) growth rate at $t\Omega_{cp} \approx 50$, obtained from the dispersion relation in Eq.~\eqref{dispersion_relation}. As both ions and electrons heat, the unstable MTSI modes shift toward larger perpendicular wavelengths. }
    \label{fig:linear_theory_t_50}
\end{figure}

\begin{figure}
    \centering
    \begin{subfigure}{0.5\textwidth}
    \centering
    \caption{$|\hat{E}_{y}(\vec{k}, t\Omega_{cp} = 50)|^2$}
    \includegraphics[width=\linewidth]{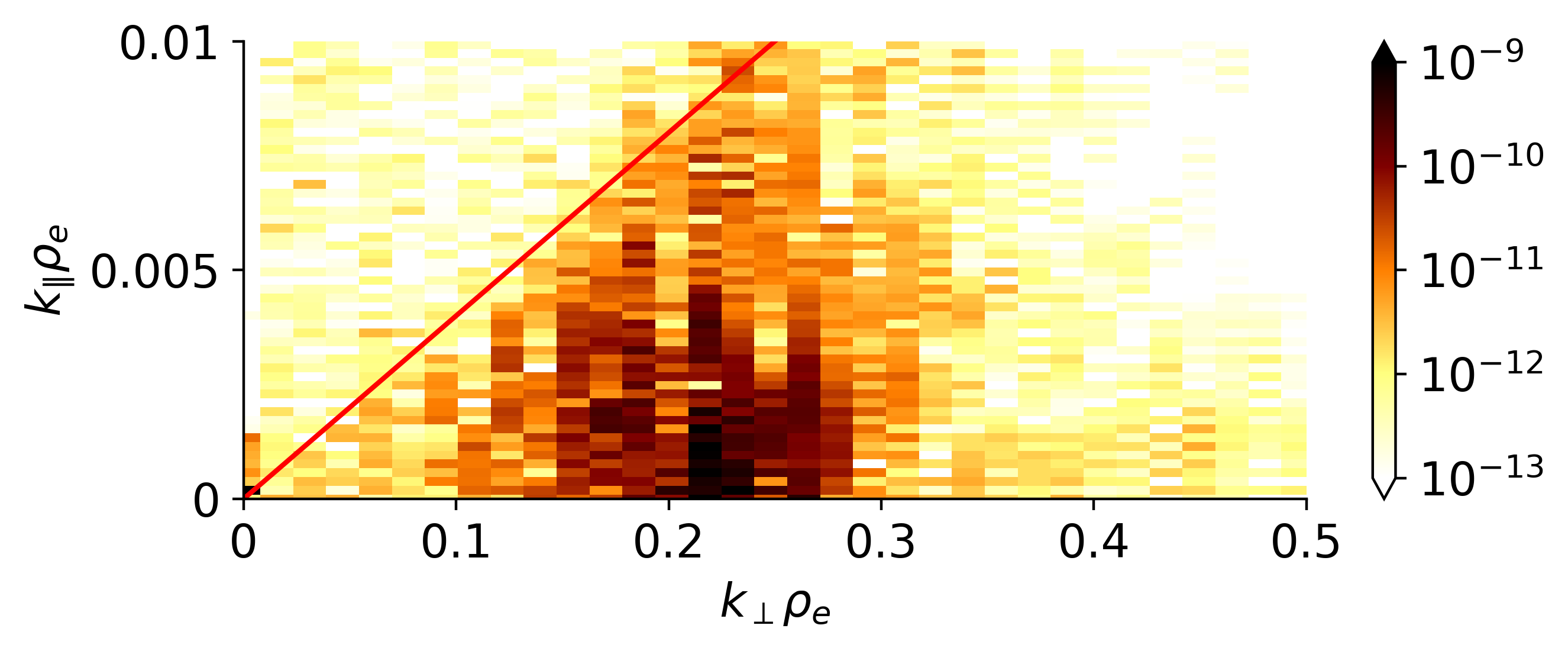}
    \label{fig:ey_fft_50}
    \end{subfigure}
    \hspace{-7pt}
    \begin{subfigure}{0.5\textwidth}
    \centering
    \caption{$|\hat{E}_{y}(\vec{k}, t\Omega_{cp} = 60)|^2$}
    \includegraphics[width=\linewidth]{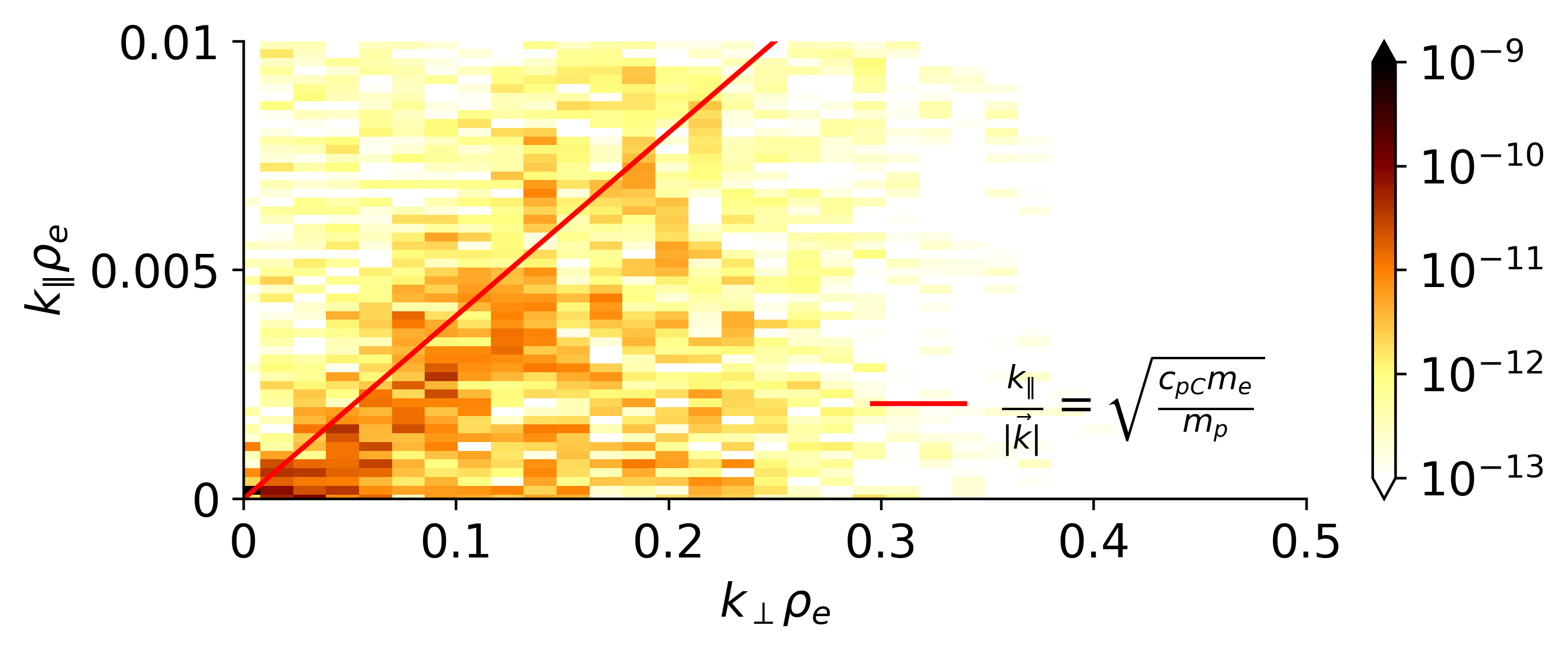}
    \label{fig:ey_fft_60}
    \end{subfigure}
    \begin{subfigure}{0.5\textwidth}
    \centering
    \caption{$n_{pC}(y, z, t\Omega_{cp} = 50)$}
    \includegraphics[width=\linewidth]{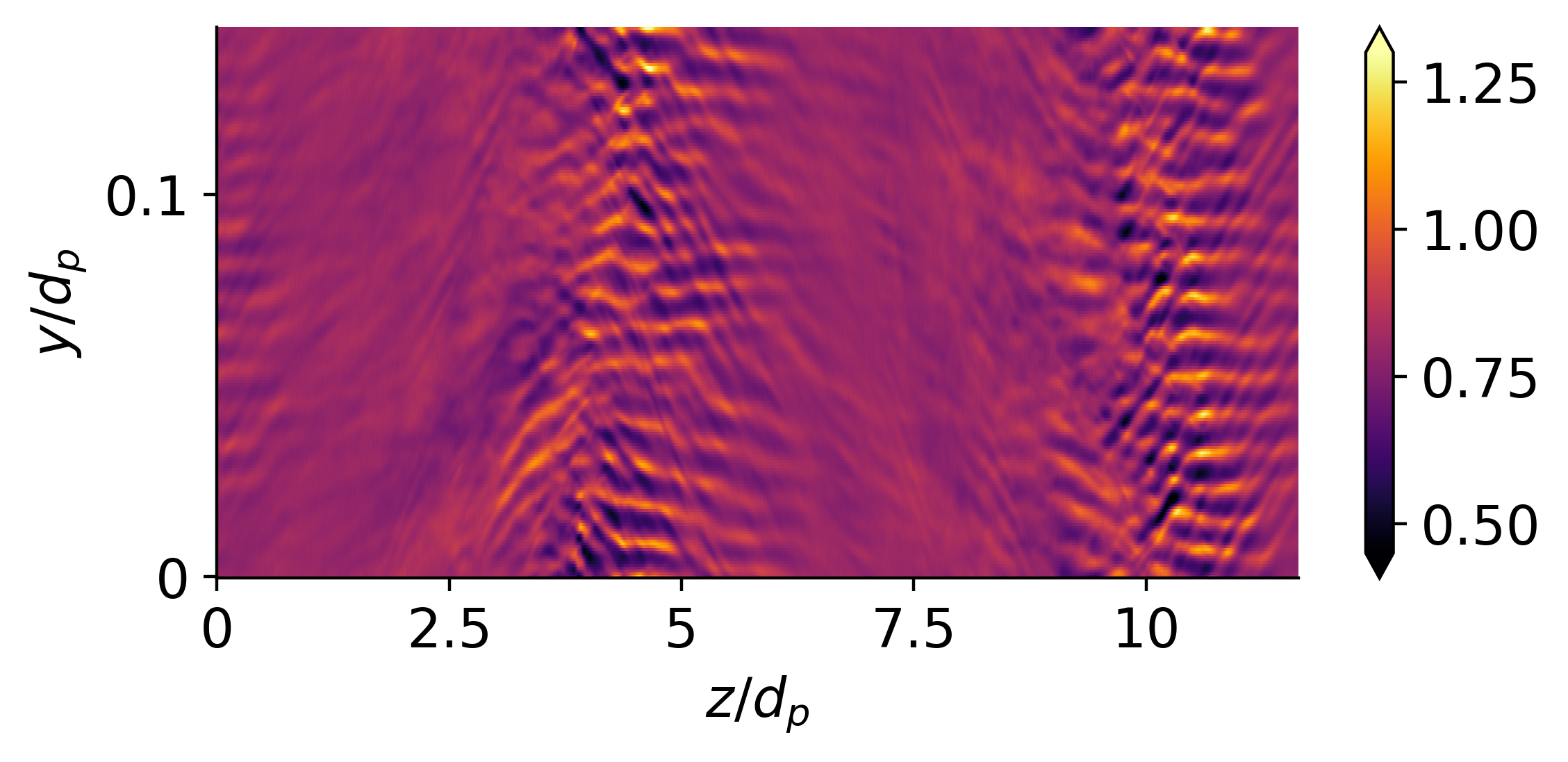}
    \end{subfigure}
    \hspace{-7pt}
    \begin{subfigure}{0.5\textwidth}
    \centering
    \caption{$n_{pC}(y, z, t\Omega_{cp} = 60)$}
    \includegraphics[width=\linewidth]{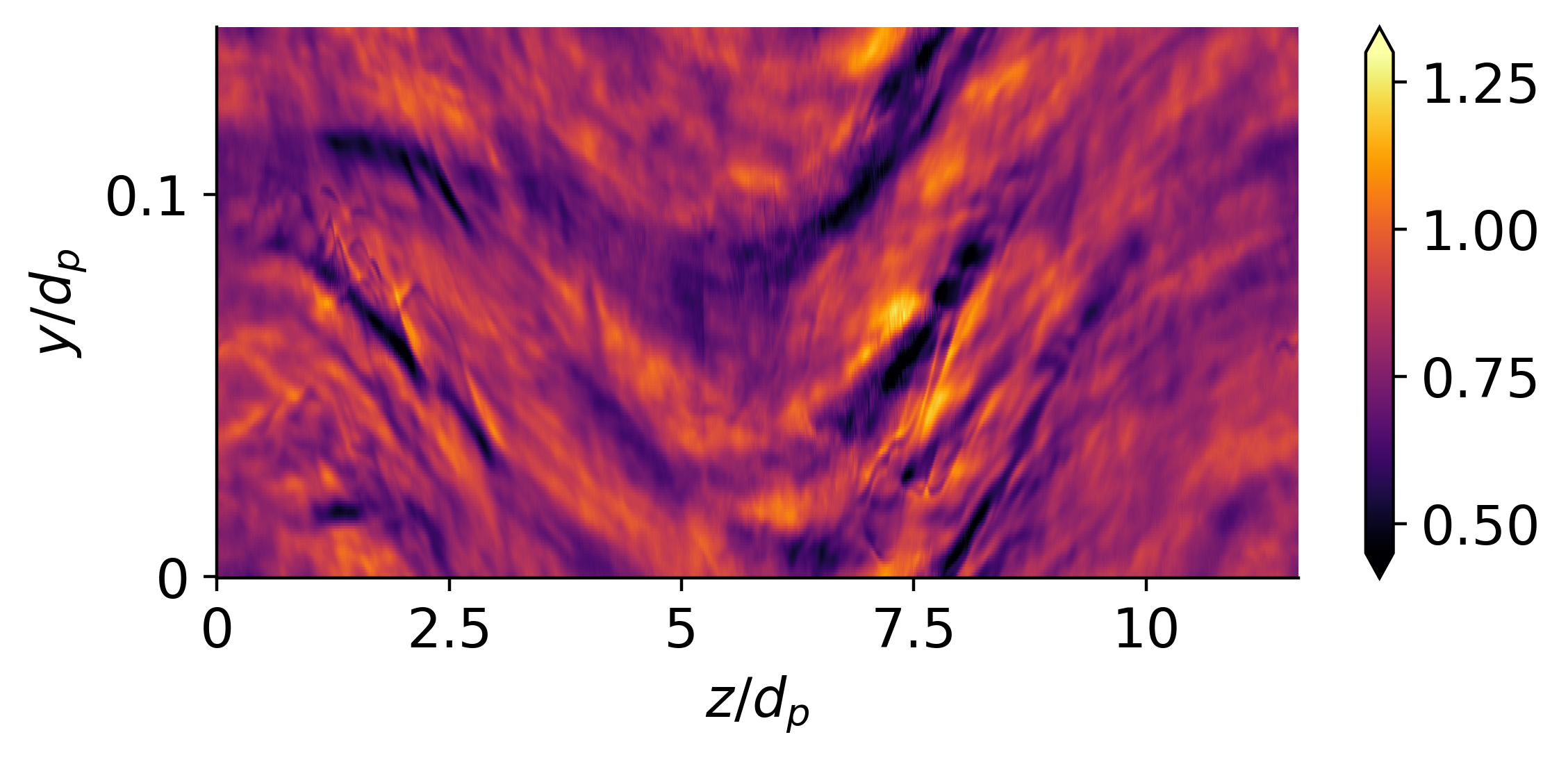}
    \end{subfigure}
    \caption{The perpendicular electric field spectrum $|\hat{E}_{y}(\vec{k}, t)|^2$ and the cold proton density $n_{pC}$ at (a/c)~$t\Omega_{cp} = 50$ and (b/d)~$t\Omega_{cp} = 60$. The dominant unstable modes transition from perpendicular ion-ion cross-field modes to oblique long-wavelength electron-proton MTSI modes.}
    \label{fig:ey_ncp_fft_2d}
\end{figure}

At $t \Omega_{cp} \approx 50$, the oxygen and cold proton populations heat rapidly in the perpendicular direction, as shown in Figure~\ref{fig:macro_quantities_tracers} panels~(d) and~(e).
Thus, their relative drift to thermal velocity ratio reduces. 
At $t\Omega_{cp} \approx 50$, the simulation parameters are $U_{yDpC} \approx U_{yDpH} \approx -2.4v_{tpC}$ and $U_{yDO+} \approx 7.9 v_{tO+}$. Relative to the PIC initialization, the electron temperature increases by factors of $1.4$ in the perpendicular direction and $1.2$ in the parallel direction. The oxygen temperature increases by a factor of $2$ in the perpendicular direction, while the cold proton perpendicular temperature increases by a factor of $3$.
The rest of the parameters are the same as in the PIC initialization, which is described in section~\ref{sec:pic_setup}.
The numerical results of the dispersion relation in Eq.~\eqref{dispersion_relation} are damped ion-ion cross-field modes and slowly growing MTSI modes.
%
Figure~\ref{fig:linear_theory_t_50} shows that the MTSI modes shift to larger perpendicular wavelength modes at $t\Omega_{cp} \approx 50$, where the most unstable mode propagates at $\theta_{k} \approx 89.5^{\circ}$ with $\gamma_{\mathrm{max}} \approx 0.004 \omega_{LH} \approx 0.08 \Omega_{cp}$ and $\omega_{r} \approx 0.04 \omega_{LH} \approx 0.8 \Omega_{cp}$. 
This corresponds to an $85\%$ reduction in the MTSI maximum growth rate compared to $t\Omega_{cp} \approx 40$.
The perpendicular electric field spectrum exhibits a similar shift in wavenumber space between $t\Omega_{cp} = 50$ and $t\Omega_{cp} = 60$ (Figure~\ref{fig:ey_ncp_fft_2d}). At $t\Omega_{cp} = 50$, the most unstable modes are associated with the perpendicular ion-ion cross-field instability, whereas following perpendicular ion heating, the dominant modes transition to oblique long wavelength electron-proton MTSI. The corresponding wavenumber ranges are consistent with the linear theory results shown in Figures~\ref{fig:linear_theory_t_40} and~\ref{fig:linear_theory_t_50}, yet the linear theory MTSI modes are less oblique than observed in the simulation. The two instabilities are clearly evident in the cold proton density shown in Figure~\ref{fig:ey_ncp_fft_2d}. At $t\Omega_{cp} \approx 50$, the fluctuations are perpendicular and localized in the $z$-direction, consistent with the ion-ion cross-field instability. By $t\Omega_{cp} \approx 60$, the fluctuations shift to longer wavelengths, extend across the full $z$-direction, and become quasi-oblique, consistent with MTSI.

\begin{figure}
    \centering
    \begin{subfigure}{0.33\textwidth}
    \centering
    \caption{$t\Omega_{cp} = 44$}
    \includegraphics[width=\linewidth]{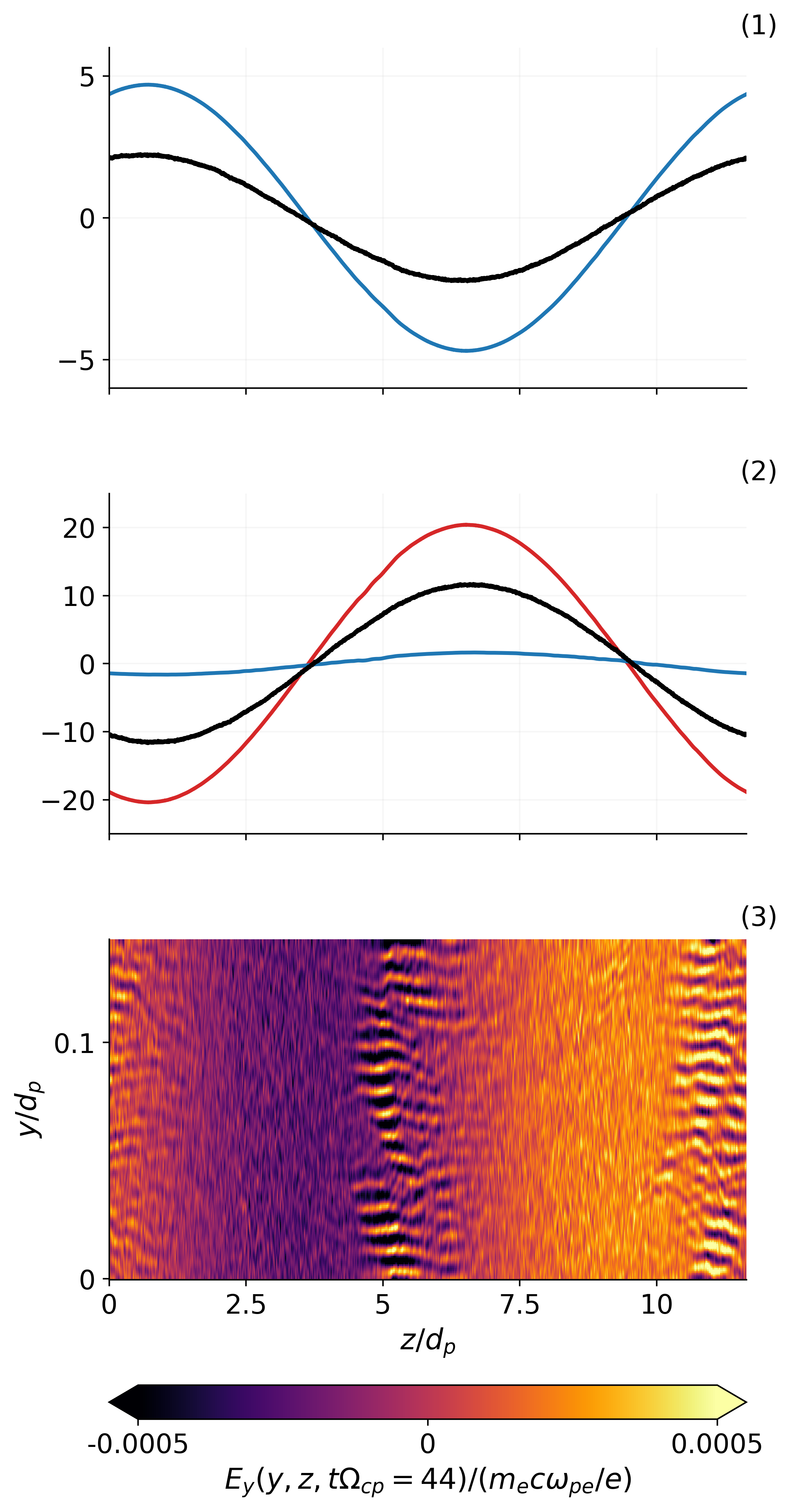}
    \end{subfigure}
    \hspace{-10pt}
    \begin{subfigure}{0.33\textwidth}
    \centering
    \caption{$t\Omega_{cp}=45$}
    \includegraphics[width=\linewidth]{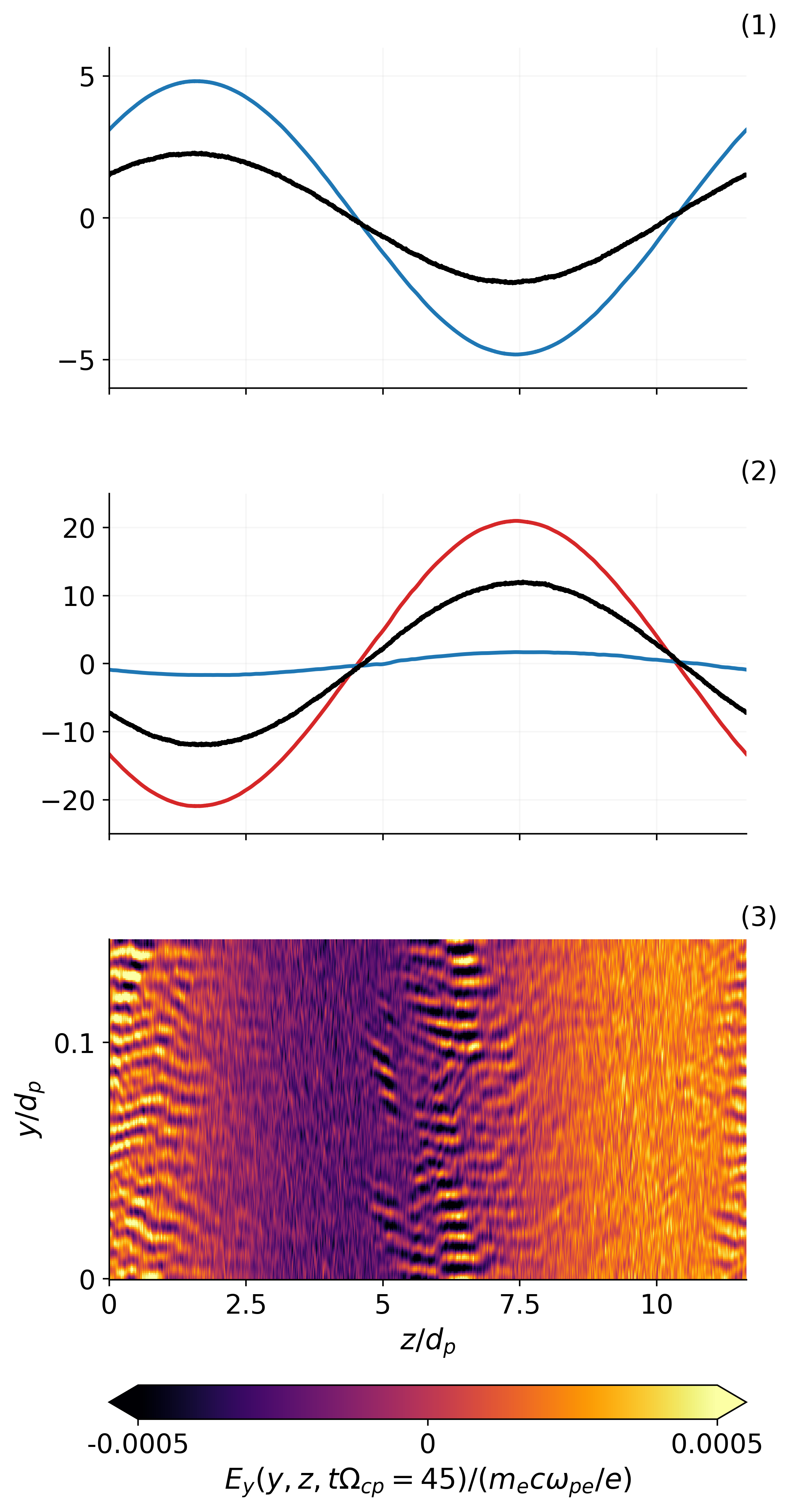}
    \end{subfigure}
    \hspace{-10pt}
    \begin{subfigure}{0.33\textwidth}
    \centering
    \caption{$t\Omega_{cp}=46$}
    \includegraphics[width=\linewidth]{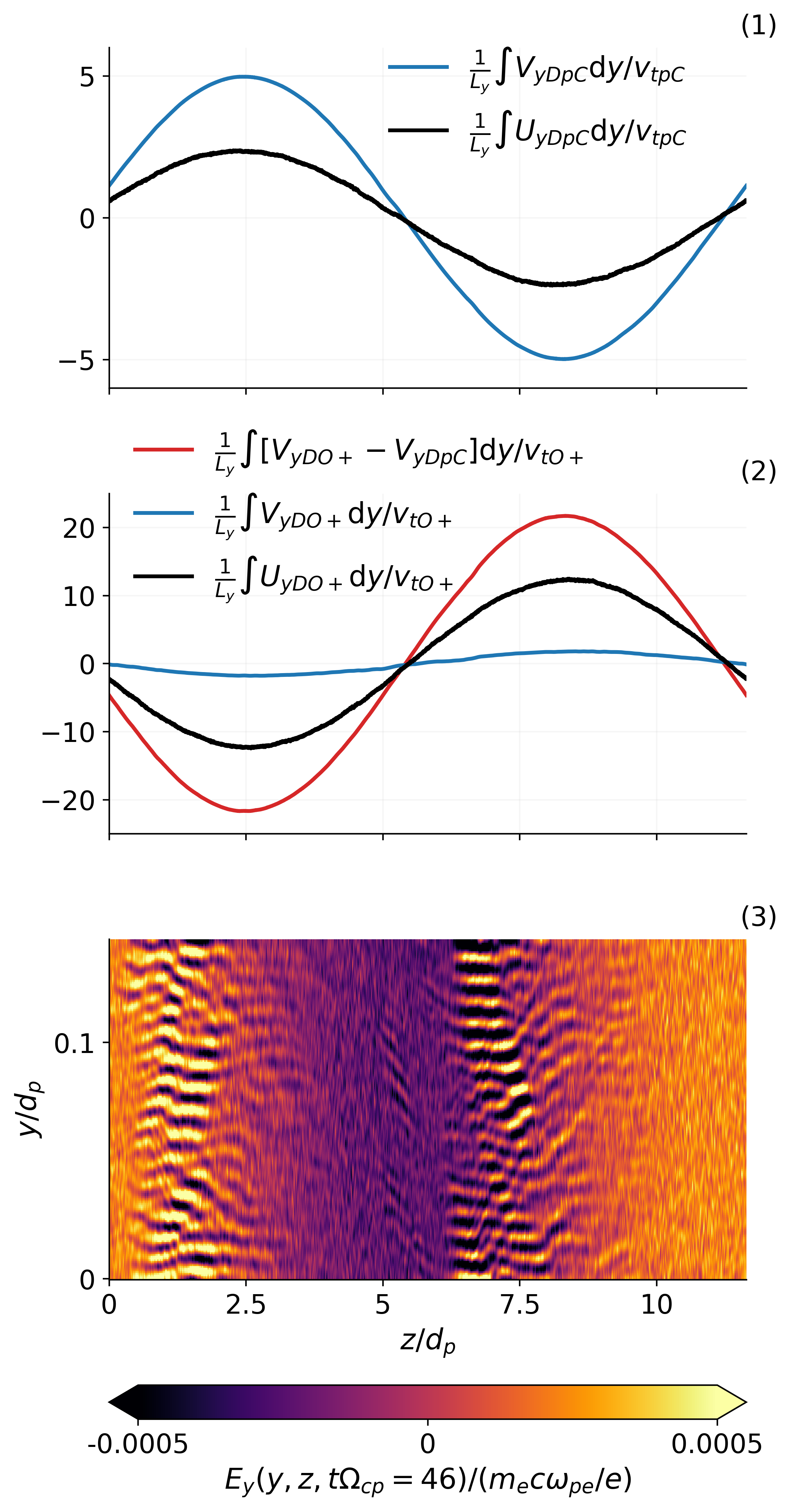}
    \end{subfigure}
    \caption{The cold populations bulk velocity and relative drift with respect to electrons at (a) $t\Omega_{cp} = 44$, (b) $t\Omega_{cp} = 45$, and (c) $t \Omega_{cp} = 46$ averaged over the $y$-direction of (1) cold protons and (2) oxygen ions. The bulk velocity is denoted by $V$ and the relative drift with respect to electrons by $U$, see~\eqref{relative_drift_y}. Additionally, the (normalized) perpendicular electric field $E_{y}(y, z, t\Omega_{cp})$ is shown in panel~(3). The secondary fluctuations are mainly perpendicular due to the ion-ion cross-field instability and are localized in $z$ depending on the location of maximum relative drift.  }
    \label{fig:drift_Ey_field_correlation}
\end{figure}

\begin{figure}
    \centering
    \begin{subfigure}{0.9\textwidth}
    \centering
    \caption{$t\Omega_{cp} = 44$ at $z=6d_{p}$}
    \includegraphics[width=\linewidth]{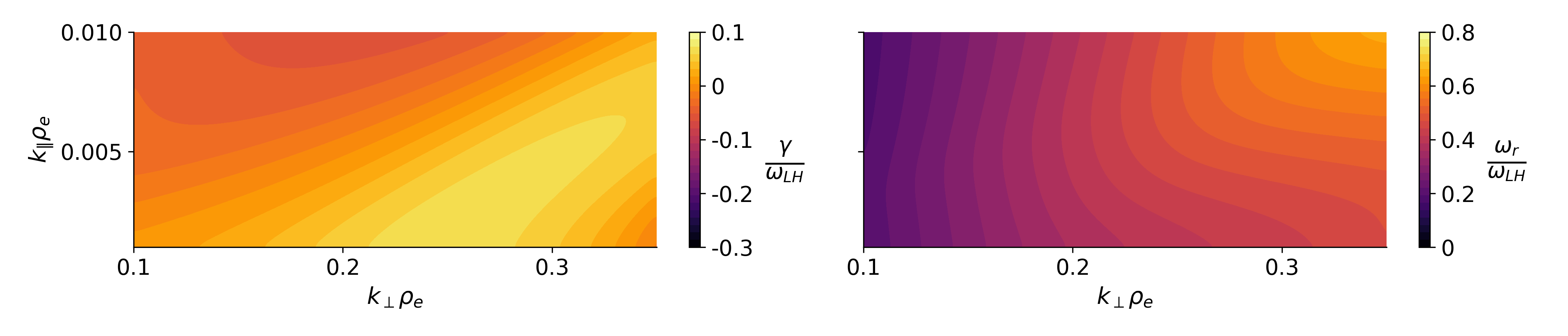}
    \end{subfigure}
    \hspace{-5pt}
    \begin{subfigure}{0.9\textwidth}
    \centering
    \caption{$t\Omega_{cp}=45$ at $z = 6d_{p}$}
    \includegraphics[width=\linewidth]{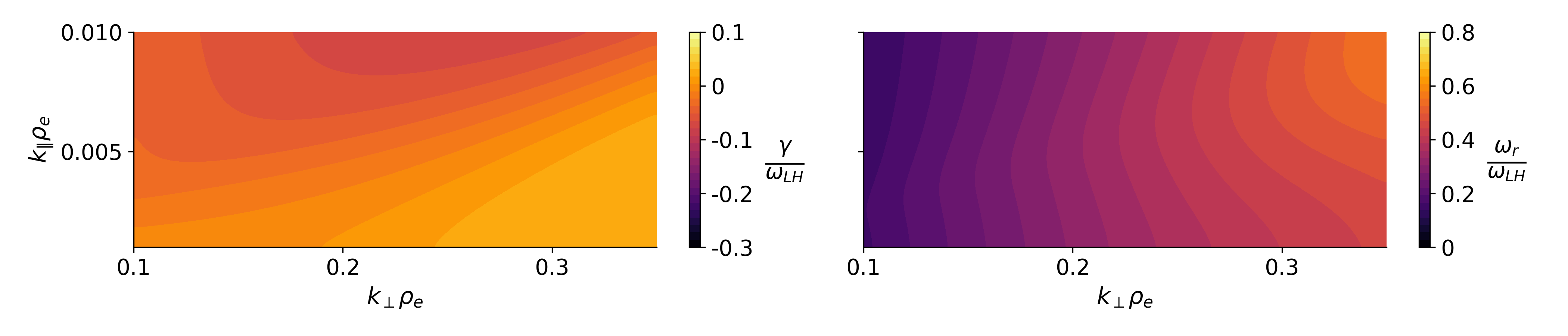}
    \end{subfigure}
    \hspace{-5pt}
    \begin{subfigure}{0.9\textwidth}
    \centering
    \caption{$t\Omega_{cp}=46$ at $z = 6d_{p}$}
    \includegraphics[width=\linewidth]{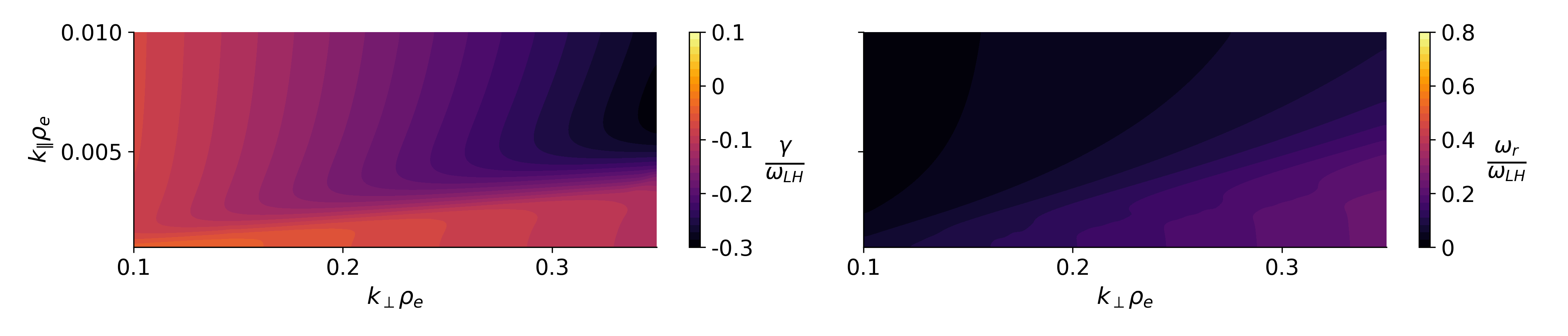}
    \end{subfigure}
    \caption{The growth rate and frequency of the ion-ion cross-field instability at $z = 6d_p$ for (a) $t\Omega_{cp} = 44$, (b) $t\Omega_{cp} = 45$, and (c) $t\Omega_{cp} = 46$. The growth rate evolves rapidly as the region of maximum relative proton-oxygen drift is advected along the parallel $z$-direction. }
    \label{fig:growth_rate_and_frequency_ion_ion}
\end{figure}

\begin{figure}
    \centering
    \begin{subfigure}{0.49\textwidth}
    \centering
    \caption{Secondary waves amplitude in time}
    \includegraphics[width=\linewidth]{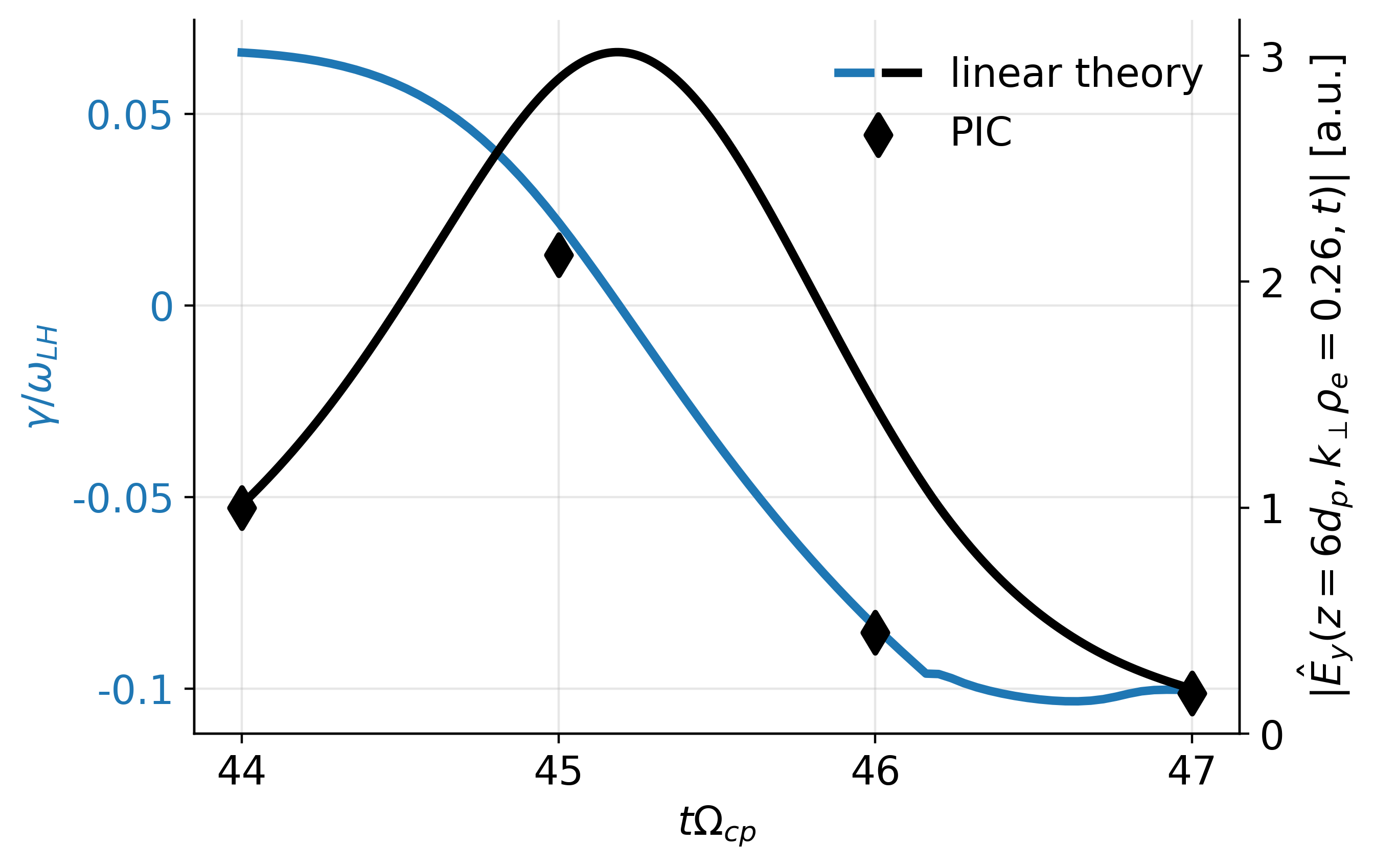}
    \label{fig:gamma_in_time_and_drift_a}
    \end{subfigure}
    \begin{subfigure}{0.49\textwidth}
    \centering
    \caption{Linear drift estimates vs. PIC}
    \includegraphics[width=\linewidth]{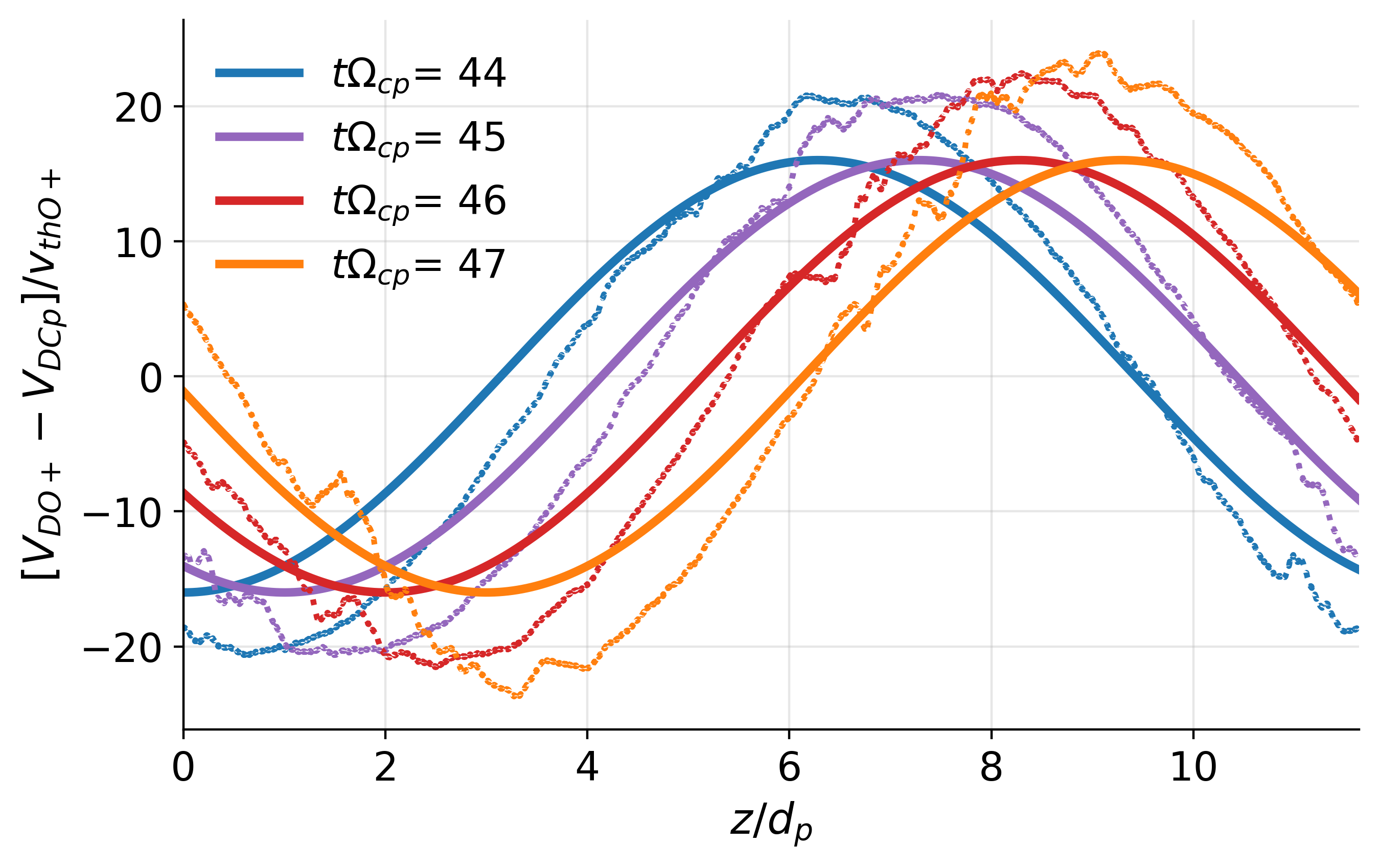}
    \label{fig:gamma_in_time_and_drift_b}
    \end{subfigure}
    \caption{The secondary ion-ion cross-field instability linear theory growth rate and amplitude are shown in subfigure~(a) and compared to the PIC simulation results at $k_{\perp} \rho_{e} = 0.26$ and $z=6d_{p}$. Subfigure~(b) shows the linear theory estimated drift in time from Eq.~\eqref{relative_drift_y} in solid lines and the PIC drifts in dotted lines, which show good agreement at $z=6d_{p}$. The linear drift estimates in subfigure~(b) are used to compute the growth rate and amplitude in subfigure~(a). }
    \label{fig:gamma_in_time_and_drift}
\end{figure}

To elucidate the spatial structure of the secondary instabilities and further compare the PIC simulation to the linear theory results, we plot the $y$-averaged bulk velocity of the cold ions along with the perpendicular electric field $E_{y}$ at $t\Omega_{cp} = \{44, 45, 46\}$ in Figure~\ref{fig:drift_Ey_field_correlation}.
For all three timestamps, the perpendicular oxygen bulk velocity (panel~(2)) is in the opposite direction to cold protons (panel~(1)). 
Also, it is evident that the electric field fluctuations are predominantly transverse, consistent with the ion-ion cross-field instability.
Since the primary wave is propagating in the positive $z$-direction rather than forming a standing wave, the localized secondary waves in Figure~\ref{fig:drift_Ey_field_correlation} emerge at different $z$ locations, determined by where the relative particle drift is largest. 
However, the peak bulk-velocity locations do not align with the fluctuations due to the reduced proton-to-electron mass ratio, $m_{p}/m_{e} = 400$, as the secondary ion-ion cross-field instability exhibits a characteristic growth time of order $\gamma^{-1} \approx 0.7\Omega_{cp}^{-1}$. 
Consequently, the fluctuations observed at time $t_{0}$ are approximately aligned with the peak bulk-velocity locations at $t_{0} - \Omega_{cp}^{-1}$. 
We study more quantitatively the time dependence of the ion-ion cross-field instability in Figure~\ref{fig:growth_rate_and_frequency_ion_ion}.
The growth rate and frequency are shown at $z=6d_{p}$ for $t\Omega_{cp} = \{44, 45, 46\}$.
We vary the relative proton-oxygen drift in time according to Eq.~\eqref{relative_drift_y}, such that the relative proton-oxygen drift decreases over time, taking the values $|U_{\perp DO+} - U_{\perp DpC}| \approx 16v_{tO+}$, $13v_{tO+}$, and $6v_{tO+}$ at $t\Omega_{cp}=44$, $45$, and $46$, respectively. 
Therefore, the growth rate decreases rapidly and eventually becomes negative.
Lastly, we track the temporal evolution of the growth rate for the perpendicular mode amplitude with $k_{\perp}\rho_{e} = 0.26$, corresponding to the maximum at $t\Omega_{cp} = 44$ in Figure~\ref{fig:gamma_in_time_and_drift_a}.
All other parameters remain fixed in time. 
The growth rate becomes negative at approximately $t\Omega_{cp} = 45.2$, coinciding with the peak of the electric field mode.
From linear theory, we also estimate the amplitude of the $y$-component of the electric field at $k_{\perp}\rho_{e} = 0.26$, assuming it scales as $\exp(\gamma(t) t)$, as shown by the solid black line.
We also plot in diamond markers the PIC $y$-component of the electric field mode amplitude at the same $k_{\perp}\rho_{e}$ as a function of time; it reaches its maximum at $t\Omega_{cp} \approx 45$, in good agreement with linear theory, and subsequently decays to approximately zero amplitude at $t\Omega_{cp} = 47$. The linear theory slightly overestimates the electric field amplitude but the timescales of the decay are comparable.
Figure~\ref{fig:gamma_in_time_and_drift_b} further shows that the estimated relative proton-oxygen drift derived from Eq.~\eqref{relative_drift_y} (solid lines) agrees well with the PIC results (dotted lines), particularly at $z = 6d_{p}$, with only a slight underestimation of the peak values.
In all, this analysis explains quantitatively why the peak of the electric field is shifted relative to the drift peak. 

\begin{figure}
    \centering
    \begin{subfigure}{0.49\textwidth}
    \centering
    \caption{Electrons}
    \includegraphics[width=\linewidth]{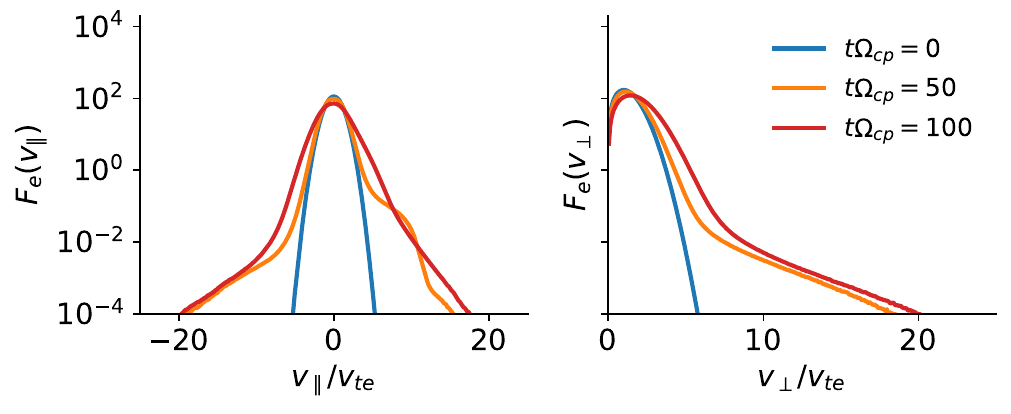}
    \end{subfigure}
        \centering
    \begin{subfigure}{0.49\textwidth}
    \centering
    \caption{Cold protons}
    \includegraphics[width=\linewidth]{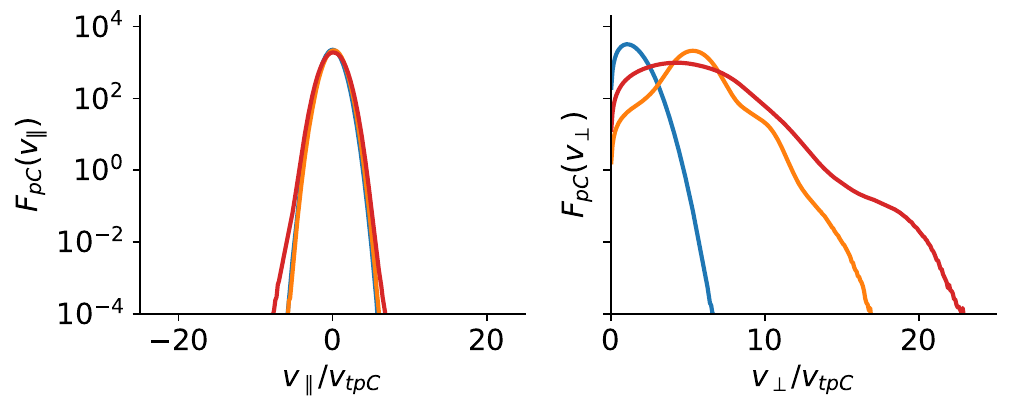}
    \end{subfigure}
    \centering
    \begin{subfigure}{0.49\textwidth}
    \centering
    \caption{Singly-charge oxygen ions}
    \includegraphics[width=\linewidth]{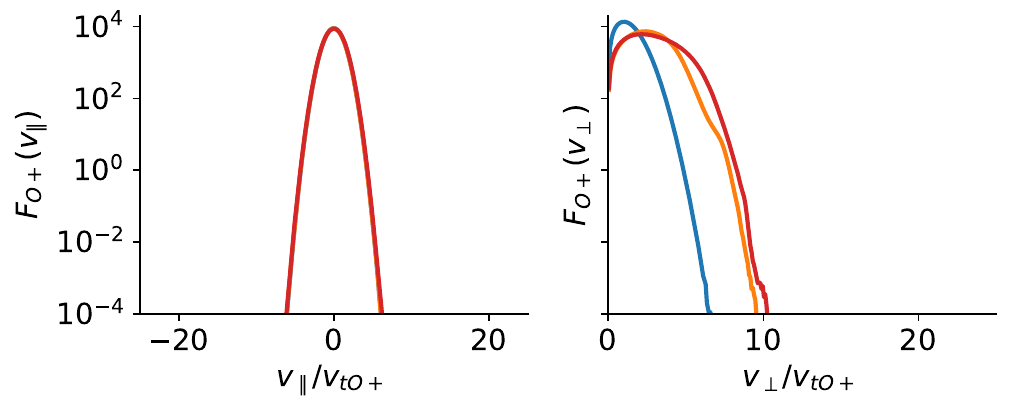}
    \end{subfigure}
\caption{Parallel and perpendicular distribution function of (a) electrons, (b) cold protons, (c) singly-charged oxygen ions in the PIC simulation.}
\label{fig:distributions_pic}
\end{figure}

Figure~\ref{fig:distributions_pic} illustrates the effects of the secondary instabilities on the distribution functions of the cold species. 
The electron distribution develops a plateau in the parallel velocity distribution at $t\Omega_{cp} \approx 50$. This is likely driven by the oblique nature of the MTSI, which leads to parallel electron heating via resonant damping, since $\omega_{r}/k_{\|} \approx 0.1 \omega_{LH} / (0.001 \rho_{e}^{-1}) \approx 5 v_{te} $.
The plateau is later removed and the distributions become smooth at later times. 
The cold protons and oxygen ions heat in the perpendicular direction likely due to nonlinear trapping as seen in previous ion-ion cross-field simulations~\cite{mcbride_1972_mtsi}.
The protons continue to heat perpendicularly after $t\Omega_{cp} \approx 50$ by the slowly growing MTSI modes. 
Moreover, the perpendicular bulk velocity of both ion populations is shifted toward higher values at $t\Omega_{cp} \approx 50$ as a result of the polarization drift induced by the EMIC electric field.

\begin{figure}
    \centering
    \begin{subfigure}{0.495\textwidth}
    \centering
    \caption{Primary instability frequency at $t\Omega_{cp} =100$}
    \includegraphics[width=\linewidth]{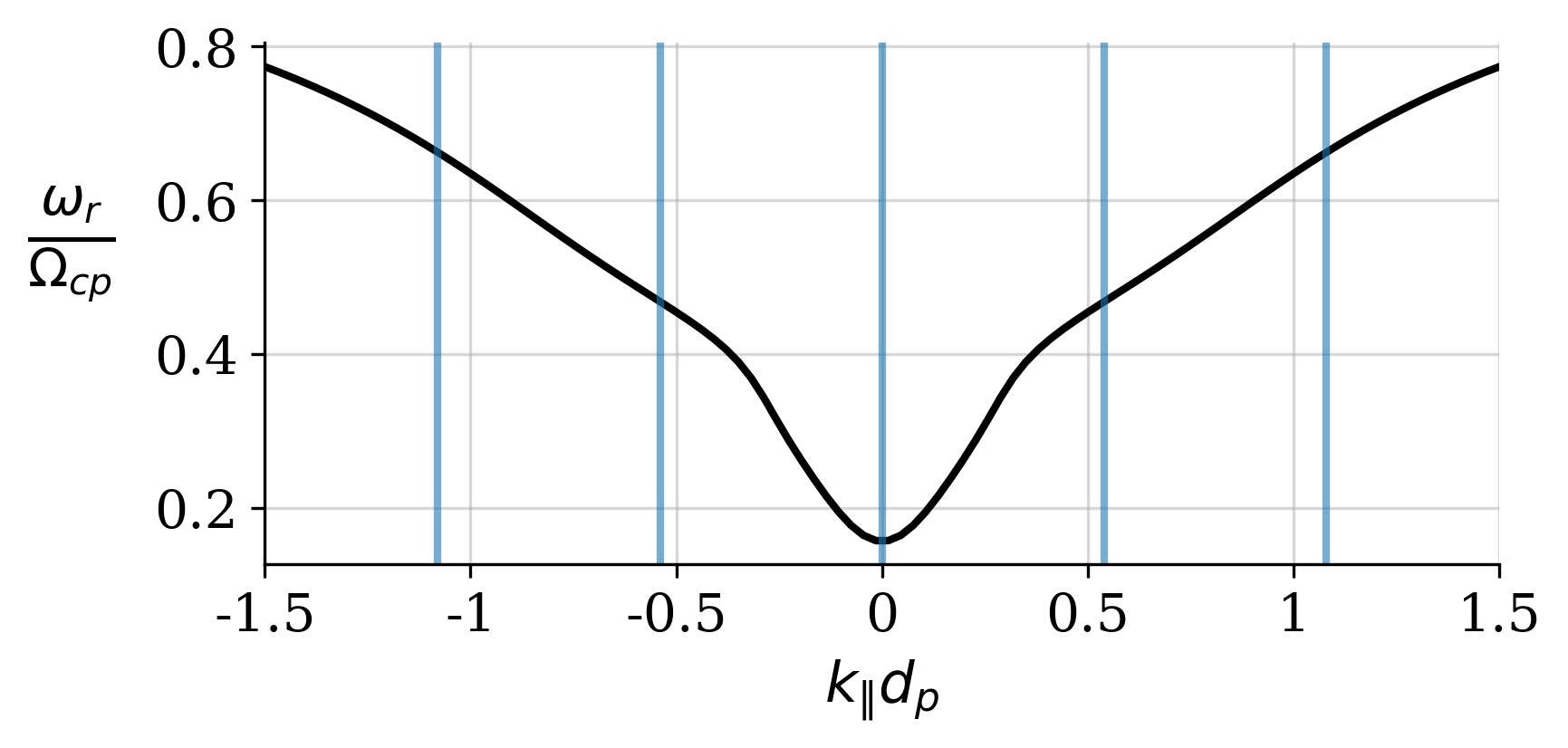}
    \end{subfigure}
    \begin{subfigure}{0.495\textwidth}
    \centering
    \caption{Primary instability growth rate at $t\Omega_{cp} = 100$}
    \includegraphics[width=\linewidth]{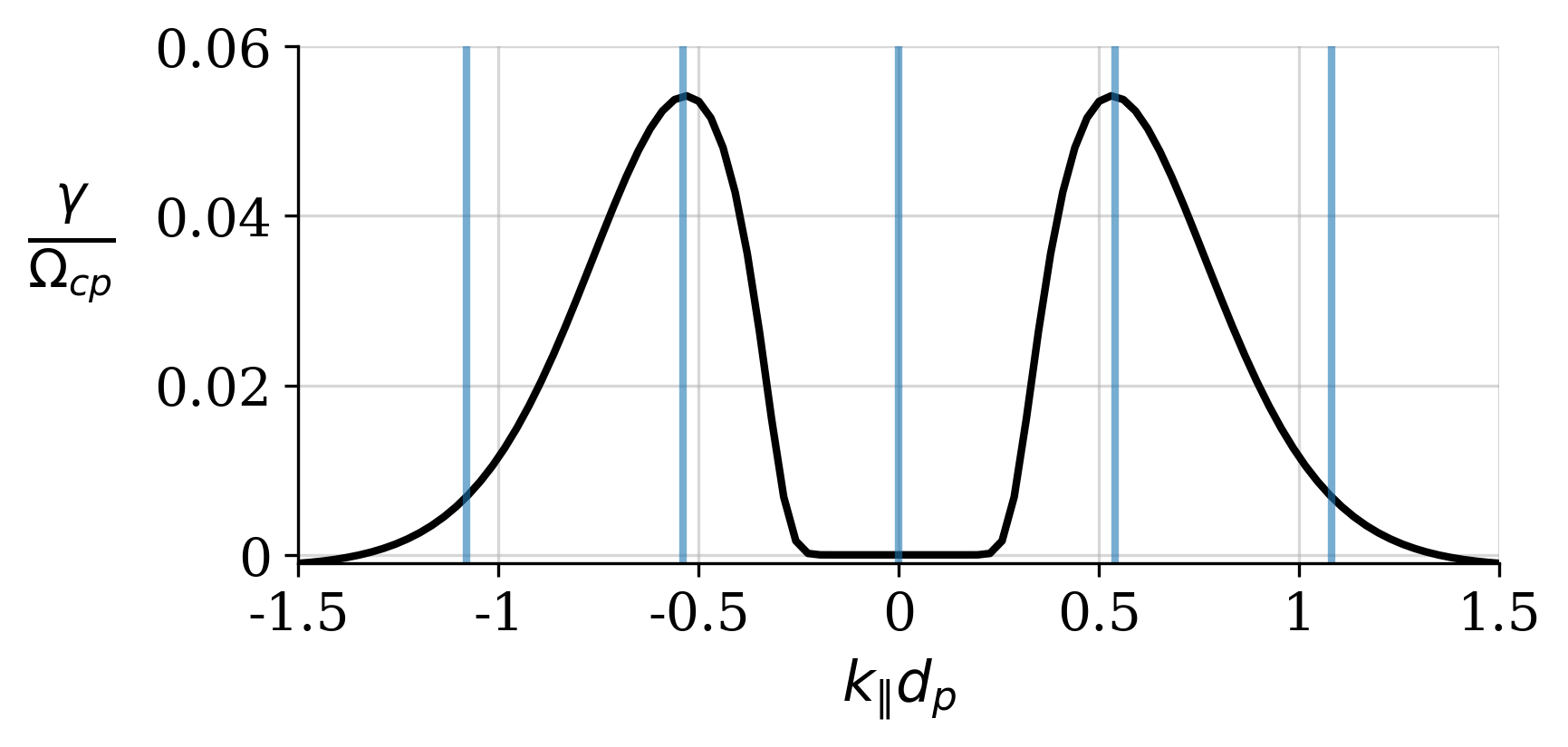}
    \end{subfigure}
    \caption{Same as Figure~\ref{fig:primary_instability_dispersion_relation}, but at $t\Omega_{cp} = 100$, where $A_{pH} = 2.8$. The maximum growth rate is still large with $\gamma_{max} \approx 0.054$.}
    \label{fig:primary_instability_dispersion_relation_100}
\end{figure}

Lastly, the dispersion relation of the primary proton cyclotron anisotropy instability at $t \Omega_{cp}= 100$ is shown in Figure~\ref{fig:primary_instability_dispersion_relation_100}. 
The parameters are $A_{pH} = 2.84, A_{pC} = 4, A_{O+} = 2, A_{e} = -0.24$, along with $\alpha_{\|e}= 0.01c$, $\alpha_{\|pC} = 2.4\times 10^{-4} c$, $\alpha_{\|pH} = 2.5\times 10^{-2}c$, $\alpha_{\|O+} = 6.2 \times 10^{-5}c$. 
The most unstable mode remains at $\omega_{0} = 0.5\Omega_{cp}$ and $k_{\|0}d_{p} \approx 0.5$.
The maximum growth rate remains substantial, with $\gamma_{\max} \approx 0.054\Omega_{cp}$, only slightly reduced from its initial value of $0.055\Omega_{cp}$ at $t = 0$ (see Figure~\ref{fig:primary_instability_dispersion_relation}).
We note that the PIC simulation perpendicular magnetic perturbations grow in a reduced rate of $\gamma \approx 0.03\Omega_{cp}$ as estimated by Figure~\ref{fig:macro_quantities_tracers} panel (a).
To conclude, although the EMIC wave saturates at $t\Omega_{cp} \approx 50$, as shown in Figure~\ref{fig:macro_quantities_tracers}, the system has not yet reached marginal stability. We suspect that the secondary instabilities lead to the early saturation of the primary wave.
Subsequently, as the secondary waves are damped, the primary instability resumes towards marginal stability, further reducing the hot proton anisotropy.

\section{Conclusions}\label{sec:conclusion}
To summarize, we present a fully kinetic simulation that quantifies the effect of secondary drift-driven instabilities generated by the electric field of a parallel-propagating EMIC wave.
We consider a multi-component plasma consisting of hot protons, cold protons, cold electrons, and cold singly-charged oxygen ions, representative of typical magnetospheric plasma composition.
The two types of secondary instabilities are the MTSI and the ion-ion cross-field instability. 
These secondary instabilities are in the lower-hybrid frequency range and occur on spatial scales comparable to the electron gyroradius.
They have previously been investigated using linear theory~\cite{khazanov_1997_lh_emic, khazanov_1996_lh_emic, sizonenko_1967_emic} and have also been observed in conjunction with EMIC waves in Earth's magnetosphere~\cite{khazanov_2017_van_allen, liu_2025_nature, pottelette_1990_jgr}.
These secondary instabilities require computationally demanding multi-dimensional simulations that resolve the characteristic scales of the cold electron population.
For the parameters considered, the ion-ion cross-field instability appears first in the simulations, which leads to predominantly perpendicular heating of cold protons and oxygen ions. 
The MTSI modes are slowly growing and lead primarily to parallel and perpendicular heating of electrons and minor perpendicular heating of cold protons. 
At the same time, the primary EMIC wave amplitude decreases by 32\%.
This process can be viewed as EMIC waves mediating the transfer of energy from hot protons to colder populations.
As the secondary modes decay, the EMIC wave continues its exponential growth, while the hot proton anisotropy continues to decrease, since the proton cyclotron anisotropy instability has not yet reached marginal stability.
This may trigger again in later times the drift-driven secondary instabilities which will continue to heat the cold populations. 
Therefore, the presence of cold ions appears to delay the time required for the EMIC wave to reach saturation. 
This may also influence estimates of EMIC driven loss processes and potential remediation strategies~\cite{desoria_2013_rbr}.

The secondary instabilities examined here lead to only moderate heating of the cold populations, raising their perpendicular temperature by $\sim$2--5 times (although only for the relatively short time scales attainable in our fully kinetic PIC simulation).
By contrast, other nonlinear processes identified in hybrid simulations produce significantly stronger heating. 
For example, the nonlinear nonresonant mechanism reported by~\citet{omidi_2010_phase_bunching} results in up to three orders of magnitude perpendicular heating for cold singly-charged helium and roughly a sixfold increase for cold protons. 
Recent hybrid simulations by~\citet{gu_2025_jgr} incorporate a minor population of cold singly-charged oxygen ions, leading to a twentyfold increase in both parallel and perpendicular energization.
However, this nonlinear mechanism operates on much longer timescales ($\sim 10^{3} - 10^{4}$ proton gyroperiods, i.e., minutes to hours), whereas the secondary instabilities considered here evolve rapidly ($\sim 50$ proton gyroperiods, i.e., seconds).
For future work, we plan to compare hybrid and fully kinetic PIC simulations for evaluating the relative importance of drift-driven instabilities versus other nonlinear processes, e.g. parametric decay~\cite{porkolab_1990_parameteric, gomberoff_1995_parametric}.
Since hybrid models treat electrons as a massless fluid, they cannot capture lower-hybrid waves, making them well suited for isolating ion-kinetic nonlinear processes and removing lower-hybrid secondary instability effects.
We also aim to clarify how the relative importance of these mechanisms depends on EMIC wave frequency and amplitude.
Finally, we plan to include a minor singly-charged helium population in future simulations to better represent magnetospheric conditions.

\section*{Acknowledgment}
O.I. was partially supported by the Los Alamos National Laboratory (LANL) Student Fellowship sponsored by the Center for Space and Earth Science (CSES). 
CSES is funded by LANL's Laboratory Directed Research and Development (LDRD) program under project number 20210528CR. 
O.I. was partially supported by the Strategic Enhancement of Excellence through Diversity Fellowship at the University of California, San Diego, in the Department of Mechanical and Aerospace Engineering. 
S.J. has been partially supported by the LANL LDRD Program under project numbers 20230341ER and 20240735DI.
The LANL LDRD Program supported G.L.D. under project number 20250577ER. LANL is operated by Triad National Security, LLC, for the National Nuclear Security Administration of the US Department of Energy (Contract No. 89233218CNA000001).
V.R. and P.K. at the Space Science Institute were partially supported by NSF award 2031024 and NASA grant 80NSSC22K1014. 
PIC simulations utilized in this work were performed using resources provided by the NASA High-End Computing (HEC) Program through the NASA Advanced Supercomputing (NAS) Division at Ames Research Center.

\appendix
\section{Derivation of the electron Vlasov equation co-drifting with $\vec{V}_{De}(z=0, t)$}\label{sec:appendix-codrift-electron}
The electrostatic magnetized electron Vlasov equation in the laboratory frame is
\begin{equation}\label{electron_vlasov_lab}
\left[\partial_{t} + \vec{v} \cdot \nabla_{\vec{x}}
-\frac{e}{m_e} \left[\vec{E}_{D}(z=0, t)  + \delta \vec{E}(\vec{x}, t) + \frac{\vec{v} \times \vec{B}_{0}}{c} \right] \cdot \nabla_{\vec{v}} \right] f_e(\vec{x}, \vec{v}, t) = 0,
\end{equation}
where $\vec{E}_{D}(z=0, t)$ is the EMIC electric field~\eqref{EMIC_E_field} and $\delta \vec{E}(\vec{x}, t)$ is the secondary waves electric field. 
Our goal is to rewrite the electron Vlasov equation in the laboratory frame~\eqref{electron_vlasov_lab} in a frame co-drifting with the electrons, $(\vec{x}, \vec{v}) \to (\vec{x}', \vec{v}')$, as defined in Eq.~\eqref{co_drifting_frame}.
Next, we apply the chain rule for time, velocity, and space derivatives:
\begin{align*}
    \partial_{t} f_e(\vec{x}, \vec{v}, t)
    &= \partial_{t} f_{e}(\vec{x}', \vec{v}', t)
    + \frac{\partial \vec{x}'}{\partial t} \cdot \nabla_{\vec{x}'} f_e(\vec{x}', \vec{v}', t)
    + \frac{\partial \vec{v}'}{\partial t} \cdot \nabla_{\vec{v}'} f_e(\vec{x}', \vec{v}', t) \\
    &= \left[ \partial_{t} - \vec{V}_{De}(z=0, t) \cdot \nabla_{\vec{x}'}  - \frac{\mathrm{d} \vec{V}_{De}(z=0, t)}{\mathrm{d} t} \cdot \nabla_{\vec{v}'} \right] f_e(\vec{x}', \vec{v}', t), \\
    \nabla_{\vec{v}} f_{e}(\vec{x}, \vec{v}, t) &= \frac{\partial \vec{v}'}{\partial \vec{v}}  \frac{\partial }{\partial \vec{v}'} f_{e}(\vec{x}', \vec{v}', t) = \nabla_{\vec{v}'} f_{e}(\vec{x}', \vec{v}', t),\\
    \nabla_{\vec{x}} f_e(\vec{x}, \vec{v}, t) &= \frac{\partial \vec{x}'}{\partial \vec{x}}  \frac{\partial }{\partial \vec{x}'} f_{e}(\vec{x}', \vec{v}', t)= 
    \nabla_{\vec{x}'} f_{e}(\vec{x}', \vec{v}', t).
\end{align*}
Therefore, substituting the above identities and Eq.~\eqref{dVdt} into Eq.~\eqref{electron_vlasov_lab} yields the electron Vlasov equation in the co-drifting frame, as given in Eq.~\eqref{vlasov_cold_co_drifting}.

\bibliography{references}

@article{gary_1976_jgr,
    title="{Proton temperature anisotropy instabilities in the solar wind}",
    author={Gary, S. P. and Montgomery, M. D. and Feldman, W. C. and Forslund, D. W.},
    journal={Journal of Geophysical Research},
    volume={81},
    number={7},
    pages={1241--1246},
    year={1976},
    publisher={Wiley Online Library}
}

@book{gary_1993_theory,
    title={Theory of Space Plasma Microinstabilities},
    author={Gary, S. P.},
    year={1993},
    publisher={Cambridge University Press}
}

@article{gary_1987_drift,
    author = {Gary, S. P. and Tokar, R. L. and Winske, D.},
    title = "{Ion/ion and electron/ion cross-field instabilities near the lower hybrid frequency}",
    journal = {Journal of Geophysical Research: Space Physics},
    volume = {92},
    number = {A9},
    pages = {10029-10038},
    year = {1987}
}

@article{kim_2025_emic_h+,
    author = {Kim, E. H. and Kim, K. H. and Johnson, J. R. and Damiano, P. A. and Shiraiwa, S. and Lin, Y. and Martin, W.},
    title = "{Prediction of the Wave Normal Angle of Proton-Band EMIC Waves Near Geosynchronous Orbit}",
    journal = {Journal of Geophysical Research: Space Physics},
    volume = {130},
    pages = {e2025JA034612},
    number = {12},
    year = {2025}
}

@article{horne_1997_emic_he+,
    title="{Wave heating of He$^{+}$ by electromagnetic ion cyclotron waves in the magnetosphere: Heating near the H$^{+}$--He$^{+}$ bi-ion resonance frequency}",
    author={Horne, R. B. and Thorne, R. M.},
    journal={Journal of Geophysical Research: Space Physics},
    volume={102},
    number={A6},
    pages={11457--11471},
    year={1997},
    publisher={Wiley Online Library}
}

@article{bashir_2021_emic_n+,
    author = {Bashir, M. F. and Ilie, R.},
    title = "{The First Observation of N$^{+}$ Electromagnetic Ion Cyclotron Waves}",
    journal = {Journal of Geophysical Research: Space Physics},
    volume = {126},
    pages = {e2020JA028716},
    number = {3},
    year = {2021}
}

@article{xiongdong_2015_emic_o+,
    author = {Yu, X. and Yuan, Z. and Wang, D. and Li, H. and Huang, S. and Wang, Z. and Zheng, Q. and Zhou, M. and Kletzing, C. A. and Wygant, J. R.},
    title = "{In situ observations of EMIC waves in O$^{+}$ band by the Van Allen Probe A}",
    journal = {Geophysical Research Letters},
    volume = {42},
    number = {5},
    pages = {1312-1317},
    year = {2015}
}

@article{kennel_1966_jgr,
    author = {Kennel, C. F. and Petschek, H. E.},
    title = "{Limit on stably trapped particle fluxes}",
    journal = {Journal of Geophysical Research (1896-1977)},
    volume = {71},
    number = {1},
    pages = {1-28},
    year = {1966}
}

@article{cornwall_1970_jgr,
    author = {Cornwall, J. M. and Coroniti, F. V. and Thorne, R. M.},
    title = "{Turbulent loss of ring current protons}",
    journal = {Journal of Geophysical Research (1896-1977)},
    volume = {75},
    number = {25},
    pages = {4699-4709},
    year = {1970}
}

@article{cuperman_1981_anisotropy_review,
    author = {Cuperman, S.},
    title = "{Electromagnetic kinetic instabilities in multicomponent space plasmas: Theoretical predictions and computer simulation experiments}",
    journal = {Reviews of Geophysics},
    volume = {19},
    number = {2},
    pages = {307-343},
    year = {1981}
}

@article{gary_1994_jgr_emic,
  title="{Hot proton anisotropies and cool proton temperatures in the outer magnetosphere}",
  author={Gary, S. P. and Moldwin, M. B. and Thomsen, M. F. and Winske, D. and McComas, D. J.},
  journal={Journal of Geophysical Research: Space Physics},
  volume={99},
  number={A12},
  pages={23603--23615},
  year={1994},
  publisher={Wiley Online Library}
}

@article{delaznno_2021_cold_impact,
    title = "{The impact of cold electrons and cold ions in magnetospheric physics}",
    journal = {Journal of Atmospheric and Solar-Terrestrial Physics},
    volume = {220},
    pages = {105599},
    year = {2021},
    author = {G. L. Delzanno and J. E. Borovsky and M. G. Henderson and P. A. {Resendiz Lira} and V. Roytershteyn and D. T. Welling}
}

@article{maldonado_2023_frontiers,
    author={Maldonado, C. A.  and Resendiz Lira, P. A.  and Delzanno, G. L.  and Larsen, B. A.  and Reisenfeld, D. B.  and Coffey, V.},
    title="{A review of instrument techniques to measure magnetospheric cold electrons and ions}",
    journal={Frontiers in Astronomy and Space Sciences}, 
    volume={9},
    year={2023}
}

@article{omidi_2010_phase_bunching,
    author = {Omidi, N. and Thorne, R. M. and Bortnik, J.},
    title = "{Nonlinear evolution of EMIC waves in a uniform magnetic field: 1. Hybrid simulations}",
    journal = {Journal of Geophysical Research: Space Physics},
    volume = {115},
    number = {A12},
    pages = {},
    year = {2010}
}

@article{omura_1985_jgr,
    title="{Heating of thermal helium in the equatorial magnetosphere: A simulation study}",
    author={Omura, Y. and Ashour-Abdalla, M. and Gendrin, R. and Quest, K.},
    journal={Journal of Geophysical Research: Space Physics},
    volume={90},
    number={A9},
    pages={8281--8292},
    year={1985},
    publisher={Wiley Online Library}
}

@article{kwon_2023_jgr,
    title="{Energization of cold protons and helium ions by EMIC waves in the inner magnetosphere: Hybrid simulations}",
    author={Kwon, J. W. and Kim, K. H. and Jin, H. and Min, K. and Lee, S. Y. and Lee, E.},
    journal={Journal of Geophysical Research: Space Physics},
    volume={128},
    number={5},
    pages={e2022JA031240},
    year={2023},
    publisher={Wiley Online Library}
}

@article{abid_2021_pop,
  title="{Energization of cold ions by electromagnetic ion cyclotron waves: Magnetospheric multiscale (MMS) observations}",
  author={Abid, A. A. and Lu, Q. and Gao, X. L. and Alotaibi, B. M. and Ali, S. and Qureshi, M. N. S. and Al-Hadeethi, Y. and Wang, S.},
  journal={Physics of Plasmas},
  volume={28},
  number={7},
  year={2021},
  publisher={AIP Publishing}
}

@article{bortnik_2010_phase_bunching,
    author = {Bortnik, J. and Thorne, R. M. and Omidi, N.},
    title = "{Nonlinear evolution of EMIC waves in a uniform magnetic field: 2. Test-particle scattering}",
    journal = {Journal of Geophysical Research: Space Physics},
    volume = {115},
    number = {A12},
    pages = {},
    year = {2010}
}

@article{mauk_1982_phase_bunching_grl,
    title="{Electromagnetic wave energization of heavy ions by the electric ``phase bunching" process}",
    author={Mauk, B. H.},
    journal={Geophysical Research Letters},
    volume={9},
    number={10},
    pages={1163--1166},
    year={1982},
    publisher={Wiley Online Library}
}

@article{berchem_1985_phase_bunching,
  title="{Nonresonant interaction of heavy ions with electromagnetic ion cyclotron waves}",
  author={Berchem, J. and Gendrin, R.},
  journal={Journal of Geophysical Research: Space Physics},
  volume={90},
  number={A11},
  pages={10945--10960},
  year={1985},
  publisher={Wiley Online Library}
}

@article{qian_1990_jgr,
  title="{Particle simulation of ion heating in the ring current}",
  author={Qian, S. and Hudson, M. K. and Roth, I.},
  journal={Journal of Geophysical Research: Space Physics},
  volume={95},
  number={A2},
  pages={1001--1013},
  year={1990},
  publisher={Wiley Online Library}
}

@article{young_1981_jgr,
    author = {Young, D. T. and Perraut, S. and Roux, A. and de Villedary, C. and Gendrin, R. and Korth, A. and Kremser, G. and Jones, D.},
    title = "{Wave-particle interactions near $\Omega_{He+}$ observed on GEOS 1 and 2 1. Propagation of ion cyclotron waves in He$^{+}$ rich plasma}",
    journal = {Journal of Geophysical Research: Space Physics},
    volume = {86},
    number = {A8},
    pages = {6755-6772},
    year = {1981}
}

@article{roux_1982_jgr,
    author = {Roux, A. and Perraut, S. and Rauch, J. L. and de Villedary, C. and Kremser, G. and Korth, A. and Young, D. T.},
    title = "{Wave-particle interactions near $\Omega_{He^{+}}$ observed on board GEOS 1 and 2: 2. Generation of ion cyclotron waves and heating of He$^{+}$ ions}",
    journal = {Journal of Geophysical Research: Space Physics},
    volume = {87},
    number = {A10},
    pages = {8174-8190},
    year = {1982}
}

@article{anderson_1994_emic_jgr,
    author = {Anderson, B. J. and Fuselier, S. A.},
    title = "{Response of thermal ions to electromagnetic ion cyclotron waves}",
    journal = {Journal of Geophysical Research: Space Physics},
    volume = {99},
    number = {A10},
    pages = {19413-19425},
    year = {1994}
}

@article{kim_2024_jgr,
    author = {Kim, K. H. and Jun, C. W. and Kwon, J. W. and Lee, J. and Shiokawa, K. and Miyoshi, Y. and Kim, E. H. and Min, K. and Seough, J. and Asamura, K. and Shinohara, I. and Matsuoka, A. and Yokota, S. and Kasahara, Y. and Kasahara, S. and Hori, T. and Keika, K. and Kumamoto, A. and Tsuchiya, F.},
    title = "{Observation and Numerical Simulation of Cold Ions Energized by EMIC Waves}",
    journal = {Journal of Geophysical Research: Space Physics},
    volume = {129},
    number = {5},
    pages = {e2023JA032361},
    year = {2024}
}

@article{gu_2025_jgr,
  title={The role of cold oxygen ions in the EMIC wave growth},
  author={Gu, S. and Cowee, M. and Fu, X. and Chen, L. and Liu, X. and Jordanova, V.},
  journal={Journal of Geophysical Research: Space Physics},
  volume={130},
  number={6},
  pages={e2024JA033661},
  year={2025},
  publisher={Wiley Online Library}
}

@article{khazanov_1997_lh_emic,
    author = {Khazanov, G. V. and Krivorutsky, E. N. and Moore, T. E. and Liemohn, M. W. and Horwitz, J. L.},
    title = "{Lower hybrid oscillations in multicomponent space plasmas subjected to ion cyclotron waves}",
    journal = {Journal of Geophysical Research: Space Physics},
    volume = {102},
    number = {A1},
    pages = {175-184},
    year = {1997}
}

@article{khazanov_2017_van_allen,
    author = {Khazanov, G. V. and Boardsen, S. and Krivorutsky, E. N. and Engebretson, M. J. and Sibeck, D. and Chen, S. and Breneman, A.},
    title = "{Lower hybrid frequency range waves generated by ion polarization drift due to electromagnetic ion cyclotron waves: Analysis of an event observed by the Van Allen Probe B}",
    journal = {Journal of Geophysical Research: Space Physics},
    volume = {122},
    number = {1},
    pages = {449-463},
    year = {2017}
}

@article{khazanov_1997_lh_observations,
    author = {Khazanov, G. V. and Liemohn, M. W. and Krivorutsky, E. N. and Horwitz, J. L.},
    title = "{A model for lower hybrid wave excitation compared with observations by Viking}",
    journal = {Geophysical Research Letters},
    volume = {24},
    number = {19},
    pages = {2399-2402},
    year = {1997}
}

@article{gamayunov_1992_alfven,
    title = "{Saturation of Alfv\'en oscillations in the ring current region due to generation of lower hybrid waves}",
    journal = {Planetary and Space Science},
    volume = {40},
    number = {4},
    pages = {477-479},
    year = {1992},
    author = {K. V. Gamayunov and E. N. Krivorutsky and A. A. Veryaev and G. V. Khazanov}
}

@article{singh_2007_jgr,
    author = {Singh, N. and Khazanov, G. and Mukhter, A.},
    title = "{Electrostatic wave generation and transverse ion acceleration by Alfv\'enic wave components of broadband extremely low frequency turbulence}",
    journal = {Journal of Geophysical Research: Space Physics},
    volume = {112},
    number = {A6},
    year = {2007}
}

@article{khazanov_2007_grl,
    author = {Khazanov, I. and Singh, N.},
    title = "{Ion and electron accelerations by large-scale shear Alfv\'en waves via cross-field instabilities}",
    journal = {Geophysical Research Letters},
    volume = {34},
    number = {20},
    pages = {},
    year = {2007}
}

@article{khazanov_1996_lh_emic,
    author = {Khazanov, G. V. and Moore, T. E. and Krivorutsky, E. N. and Horwitz, J. L. and Liemohn, M. W.},
    title = "{Lower hybrid turbulence and ponderomotive force effects in space plasmas subjected to large-amplitude low-frequency waves}",
    journal = {Geophysical Research Letters},
    volume = {23},
    number = {8},
    pages = {797-800},
    year = {1996}
}

@article{pottelette_1990_jgr,
    title="{High-frequency waves in the cusp/cleft regions}",
    author={Pottelette, R and Malingre, M and Dubouloz, N and Aparicio, B and Lundin, R and Holmgren, G and Marklund, G},
    journal={Journal of Geophysical Research: Space Physics},
    volume={95},
    number={A5},
    pages={5957--5971},
    year={1990},
    publisher={Wiley Online Library}
}

@article{liu_2025_nature,
    title="{Direct observations of cross-scale energy transfer driven by multiple-ion interactions in space plasmas}",
    author={Liu, Z. Y. and Zong, Q. G. and Wang, S. and Zhou, X. Z. and Yue, C.},
    volume={16},
    pages={11516},
    journal={Nature Communications},
    year={2025},
    publisher={Nature Publishing Group UK London}
}

@article{saito_2015_pop,
    title="{Nonlinear damping of a finite amplitude whistler wave due to modified two stream instability}",
    author={Saito, S. and Nariyuki, Y. and Umeda, T.},
    journal={Physics of Plasmas},
    volume={22},
    number={7},
    year={2015},
    publisher={AIP Publishing}
}

@article{sizonenko_1967_emic,
  title="{Plasma instability in the electric field of an ion-cyclotron wave}",
  author={Sizonenko, V. L. and Stepanov, K. N.},
  journal={Nuclear Fusion},
  volume={7},
  number={2-3},
  pages={131},
  year={1967},
  publisher={IOP Publishing}
}

@article{buneman_1962_drift,
    year = {1962},
    volume = {4},
    number = {2},
    pages = {111},
    author = {O. Buneman},
    title = "{Instability of electrons drifting through ions across a magnetic field}",
    journal = {Journal of Nuclear Energy. Part C, Plasma Physics, Accelerators, Thermonuclear Research}
}

@article{akhiezer_1975_book,
    title="{Plasma electrodynamics. Volume 1-Linear theory.}",
    author={Akhiezer, A. I. and Akhiezer, I. A. and Polovin, R. V. and Sitenko, A. G. and Stepanov, K.N.},
    journal={Oxford Pergamon Press International Series on Natural Philosophy},
    volume={1},
    year={1975}
}

@article{ott_1972_mtsi,
  title = "{Turbulent Heating in Computer Simulations of the Modified Plasma Two-Stream Instability}",
  author = {Ott, E. and McBride, J. B. and Orens, J. H. and Boris, J. P.},
  journal = {Phys. Rev. Lett.},
  volume = {28},
  issue = {2},
  pages = {88--91},
  year = {1972},
  publisher = {American Physical Society}
}

@article{papadopoulos_1971_pof,
    author = {Papadopoulos, K. and Davidson, R. C. and Dawson, J. M. and Haber, I. and Hammer, D. A. and Krall, N. A. and Shanny, R.},
    title = "{Heating of Counterstreaming Ion Beams in an External Magnetic Field}",
    journal = {The Physics of Fluids},
    volume = {14},
    number = {4},
    pages = {849-857},
    year = {1971},
    month = {04}
}

@article{barbosa_1986_pof,
    author = {Barbosa, D. D.},
    title = "{Generation of lower-hybrid noise by superthermal cross-field ion currents}",
    journal = {The Physics of Fluids},
    volume = {29},
    number = {3},
    pages = {888-890},
    year = {1986},
    month = {03}
}

@article{krall_1971_mtsi,
  title = "{Low-Frequency Instabilities in Magnetic Pulses}",
  author = {Krall, N. A. and Liewer, P. C.},
  journal = {Phys. Rev. A},
  volume = {4},
  issue = {5},
  pages = {2094--2103},
  numpages = {0},
  year = {1971},
  publisher = {American Physical Society}
}

@article{mcbride_1972_mtsi,
    author = {McBride, J. B. and Ott, E. and Boris, J. P. and Orens, J. H.},
    title = "{Theory and Simulation of Turbulent Heating by the Modified Two-Stream Instability}",
    journal = {The Physics of Fluids},
    volume = {15},
    number = {12},
    pages = {2367-2383},
    year = {1972}
}

@article{janhunen_2018_2Decdi,
    author = {Janhunen, S. and Smolyakov, A. and Sydorenko, D. and Jimenez, M. and Kaganovich, I. and Raitses, Y.},
    title = "{Evolution of the electron cyclotron drift instability in two-dimensions}",
    journal = {Physics of Plasmas},
    volume = {25},
    number = {8},
    pages = {082308},
    year = {2018},
    month = {08}
}

@book{fried_conte_1961,
    title="{The Plasma Dispersion Function; the Hilbert Transform of the Gaussian}",
    publisher = {Academic Press},
    pages = {2-3},
    year = {1961},
    author = {Fried, B. D. and Conte, S. D.}
}

@article{hunana_2019_closure, 
    title="{An introductory guide to fluid models with anisotropic temperatures. Part 2. Kinetic theory, Pad\'e approximants and Landau fluid closures}", 
    volume={85}, 
    number={6}, 
    journal={Journal of Plasma Physics}, 
    author={Hunana, P. and Tenerani, A. and Zank, G. P. and Goldstein, M. L. and Webb, G. M. and Khomenko, E. and Collados, M. and Cally, P. S. and Adhikari, L. and Velli, M.}, 
    year={2019}, 
    pages={205850603}
}

@article{bowers_2008_VPIC,
    author = {Bowers, K. J. and Albright, B. J. and Yin, L. and Bergen, B. and Kwan, T. J. T.},
    title = "{Ultrahigh performance three-dimensional electromagnetic relativistic kinetic plasma simulation}",
    journal = {Physics of Plasmas},
    volume = {15},
    number = {5},
    pages = {055703},
    year = {2008}
}

@article{porkolab_1990_parameteric,
    title = "{Invited paper: Parametric instabilities in the tokamak edge plasma in the ion cyclotron heating regimes}",
    journal = {Fusion Engineering and Design},
    volume = {12},
    number = {1},
    pages = {93-103},
    year = {1990},
    author = {M. Porkolab}
}

@article{gomberoff_1995_parametric,
  title=
  "{Parametric decays of electromagnetic ion cyclotron waves in a H$^{+}$-He$^{+}$-O$^{+}$ magnetospheric like plasma}",
  author={Gomberoff, L. and Gnavi, G. and Gratton, F. T.},
  journal={Journal of Geophysical Research: Space Physics},
  volume={100},
  number={A9},
  pages={17221--17229},
  year={1995},
  publisher={Wiley Online Library}
}

@article{desoria_2013_rbr,
  title="{Electromagnetic ion cyclotron waves for radiation belt remediation applications}",
  author={de Soria-Santacruz, M. and Martinez-Sanchez, M.},
  journal={IEEE Transactions on Plasma Science},
  volume={41},
  number={12},
  pages={3329--3337},
  year={2013},
  publisher={IEEE}
}

@article{usanova_2014_emic,
    author = {Usanova, M. E. and Drozdov, A. and Orlova, K. and Mann, I. R. and Shprits, Y. and Robertson, M. T. and Turner, D. L. and Milling, D. K. and Kale, A. and Baker, D. N. and Thaller, S. A. and Reeves, G. D. and Spence, H. E. and Kletzing, C. and Wygant, J.},
    title = "{Effect of EMIC waves on relativistic and ultrarelativistic electron populations: Ground-based and Van Allen Probes observations}",
    journal = {Geophysical Research Letters},
    volume = {41},
    number = {5},
    pages = {1375-1381},
    year = {2014}
}

@article{summers_2003_emic,
  title="{Relativistic electron pitch-angle scattering by electromagnetic ion cyclotron waves during geomagnetic storms}",
  author={Summers, D. and Thorne, R. M.},
  journal={Journal of Geophysical Research: Space Physics},
  volume={108},
  number={A4},
  year={2003},
  publisher={Wiley Online Library}
}

@article{roytershteyn_2021_pop,
    author = {Roytershteyn, V. and Delzanno, G. L.},
    title = "{Nonlinear coupling of whistler waves to oblique electrostatic turbulence enabled by cold plasma}",
    journal = {Physics of Plasmas},
    volume = {28},
    number = {4},
    pages = {042903},
    year = {2021}
}

@article{roytershteyn_2024_frontiers,
    author={Roytershteyn, V.  and Delzanno, G. L.  and Holmes, J. },
    title="{Oblique Instability of Quasi-Parallel Whistler Waves in the Presence of Cold and Warm Electron Populations}",
    journal={Frontiers in Astronomy and Space Sciences},
    volume={11},
    year={2024}
}

@article{issan_2026_pop,
    author = {Issan, O. and Roytershteyn, V. and Delzanno, G. L. and Janhunen, S.},
    title = "{Understanding cold electron impact on parallel-propagating whistler chorus waves via moment-based quasilinear theory}",
    journal = {Physics of Plasmas},
    volume = {33},
    number = {3},
    pages = {032110},
    year = {2026}
}

\end{document}